\documentclass[11pt]{article}
\usepackage{amsfonts, amsmath,amssymb,bm,enumerate, graphicx,color}
\usepackage{subfigure,algorithm,algorithmic}
\usepackage{epstopdf}
\usepackage[dvipsnames]{xcolor}
\usepackage{multirow}
\usepackage{subfigure}
\usepackage[colorlinks=true,urlcolor=blue,citecolor=purple,linkcolor=blue,bookmarks=true]{hyperref}

\allowdisplaybreaks

\def\qed{\hfill \vrule height 6pt width 6pt depth 0pt}

\newtheorem{The}{Theorem}[section]
\newtheorem{Pro}{Proposition}[section]
\newtheorem{Lem}[The]{Lemma}
\newtheorem{Def}[The]{Definition}

\newtheorem{Exa}[The]{Example}
\def\qed{\hfill \vrule height 6pt width 6pt depth 0pt}

\newcommand{\ignore}[1]{}
\newcommand{\dif}{\,\mathrm{d}}

\definecolor{grass}{rgb}{0.0, 0.5, 0.0}

\newcounter{quest}

\begin{document}

\title{Should the Ransomware be Paid?}

\author{Rui  Fang,~
        Maochao  Xu,~
        and~Peng Zhao
\thanks{R. Fang is with the Department of Mathematics, Shantou University,   China. Email: xmufr1987@hotmail.com}
\thanks{M. Xu is with the Department of Mathematics, Illinois State University, USA. Email: mxu2@ilstu.edu}
\thanks{P. Zhao is with  the School of Mathematics and Statistics, Jiangsu Normal University, China. Email:zhaop@jsnu.edu.cn}
 }

\maketitle

\begin{abstract}
Ransomware has emerged as one of the most concerned cyber risks in recent years, which has caused millions of dollars monetary loss over the world.  It typically demands a certain amount of ransom payment within a limited timeframe to decrypt the encrypted victim's files. This paper explores whether the ransomware should be paid in a novel game-theoretic model from the perspective of Bayesian game.  In particular, the new model analyzes the ransom payment strategies within the framework of incomplete information for both  hacker and victim. Our results show that there exist pure and randomized Bayesian Nash equilibria under some mild conditions for the hacker and victim. The sufficient conditions that when the ransom should be paid are presented when an organization is compromised by the ransomware attack. We further study how the costs and probabilities of cracking or recovering affect the expected payoffs of the hacker and the victim in the equilibria.  In particular, it is found that the backup option for computer files is not always beneficial, which actually depends on the related cost. Moreover, it is discovered that fake ransomware may be more than expected because of the potential high payoffs. Numerical examples are also presented for illustration.
\end{abstract}



\section{Introduction}

Ransomware, a malware invading computers, attempting to encrypt the files and
demand the ransom from the victims, has risen to an emerging cyber risk in recent years \cite{kharraz2015cutting,ali2017ransomware}. The ransomware attack is typically launched by installing the crypto-virus on the computer without the knowledge or intention of the user. When the attack is initialized, the ransomeware encrypts files or even completely locks the computer system. Then, the user will receive a message on the screen about regaining the access to the files, which typically asks the victim to pay the ransom via Bitcoin or other crypto currencies \cite{spagnuolo2014bitiodine}.   The possible damage caused  by the ransomware attack including loss of files, exposing private information, lost revenue due to loss of data or repair down-of-service, legal expenses for damage to the third party, notification of potentially affected customers, and regulatory fines and penalties. It was reported that between 2015 and 2017 the financial cost of ransomware increased dramatically from \$325 million to \$5 billion, and it is projected that the ransomware cost will reach \$11.5 billion in 2019 \cite{fred2018}.

Ransomware has been recognized theoretically and practically for over two decades \cite{ali2017ransomware}. For the payment of ransomware, there have been a lot of debates. Generally,  government agencies such as FBI do not support to pay any ransom demanded by the hacker. However, the evidence from the business leaders in US shows that
70\% of them paid the ransomware in order to resolve the incidents \cite{ibm2016}. In theory, the backup is an efficient way to prevent the ransomware. But the empirical evidence shows that only a small portion of companies (or users) regularly backup the data due to the  cost and other related factors \cite{Kabooza2009, laszka2017economics}. Therefore, whether the ransomware should be paid is still open for debating.

In academia, there exists a rich literature on the analysis of ransomware but mainly from the security  perspectives such as prevention, detection, analysis, and prediction \cite{al2018ransomware}. The game theory \cite{myerson2013game}, a  natural tool to study the interactions among entities with contrasting interests, has become popular in the area of information theory such as wireless communication network \cite{inaltekin2008analysis}, noncooperative resource allocation policies \cite{buzzi2010transmitter}, joint optimization of the sensing and transmission strategies \cite{scutari2013joint}, and fair resource allocation \cite{fossati2018fair}. However, strategic comprehension on the conflicting situation between hackers and victims is still lacking. There are only a few studies attempting to enhance the understanding of this conflicting situation via the game-theoretic approaches. 
For example, Laszka et al.  \cite{laszka2017economics} proposed a multi-stage and multi-defender security game to investigate the role of backup investments in the adversarial interaction between the hacker and the victim. Specifically,  one single hacker and two types of victims are assumed in the model. Different types of victims have different backup frequencies as their strategies, and the hacker attacks different victims with different levels of efforts. A sufficient and necessary condition on the existence of Nash equilibria is provided.  It is also illustrated that when the backup cost is relatively low, the hacker would not deploy the ransomware because the gain from ransoms would be smaller than the expenses; when the attacking cost is low, the victim could be better off by investing more in backups.  Cartwright et al. \cite{chs2018bitcoin} studied a game model from the perspective of kidnapping on the conflicting situation incurred by ransomware, and derived Nash equilibria under different settings. One particular feature of the model is that the victim is allowed to make a counter-offer to the ransom demand. In addition, the hacker can be caught by the authorities with some positive probability. More recently, Caporusso et al. \cite{caporusso2018game} proposed a game-theoretical model to formulate the post-attack situation faced by the victim and the hacker. In this model, the game starts when the hacker locks down the victim's files and demands for ransom payment. The victim has to choose between paying or not paying, and then the hacker will decide to release the decryption information or delete the files. Similar to that in \cite{laszka2017economics}, Caporusso et al.  \cite{caporusso2018game} assumed that the hacker may raise or lower the ransom amount depending on whether the victim pays the ransom. In particular, the hacker's reputation is taken into consideration. Under this setup, it is found that for the hacker, the strategy of releasing the decryption information is dominated by the option of deleting the files. However, with an affordable amount of ransom, the victim would still choose to pay the ransom.

It should be noted that all the existing studies assuming the games with complete information. However, in reality, the ransom situation can rarely be modeled as a game with complete information. Although \cite{chs2018bitcoin} touched a little bit on the situation that the hacker is uncertain regarding the victim's willingness to pay, a formal analysis from the aspect of incomplete information is missing. {\em This paper aims to fill this gap by developing a novel game-theoretical framework with incomplete information for studying the conflicting situation of ransomware}. In particular, we assume that the game starts at the time when the hacker plans the attack, which is different from the post-attack model in \cite{caporusso2018game}.
By considering the backup situation in practice, we introduce two
Bayesian games: i) in the first game, the victim does not have the backup; ii) in the second game, the victim has the recovering from the backup as one strategy. By modeling the valuation of victim's targeted files and  hacker's types as random variables (i.e., uncertainties are allowed), we first describe
the best strategy for a victim with certain valuation on the encrypted files under a ransomware situation. The pure and randomized Bayesian Nash equilibria of the two games are presented.
We further study how different exogenous parameters affect the hacker' expected payoffs, and whether the backup option would have a positive effect on mitigating the ransomware risk.

The rest of the paper is organized as follows. In Section \ref{sec:model}, we introduce the notations and describe the proposed model. Section \ref{sec:equi} discusses the Bayesian Nash equilibria of the proposed games.  In Section \ref{sec:exm}, we study the effects of different parameters on the equilibria of both the hacker and the victim. The last Section concludes our main results, and present some discussion.
%

\section{Model overview and background} \label{sec:model}
To formalize the conflicting situation between the hacker and the victim, it is natural to characterize them as two players in a game theory model. The proposed game has the following key characteristics:
\begin{itemize}
  \item {\em Uncertainty.} Either player is uncertain about some of the other's characteristics. For example, when deciding whether to pay the ransom or not, the victim is not sure what the hacker will do after getting the ransom; when encrypting the victim's files, the hacker is unable to know how exactly  the victim values the files.
  \item {\em Two-stage.} There are two clear stages as the situation proceeds. The hacker first encrypts the victim's files and asks for the ransom payment. After receiving the ransom payment demand, the victim decides to pay the ransom or not. Therefore, a two-player two-stage {\em Bayesian game} is needed to take account of the randomness arising in the conflicting situation.
  \item {\em Non-zero-sum.} The game is not necessarily a zero-sum game  due to the fact that the payoffs of the hacker and the victim are not opposite to each other.
\end{itemize}
In the following, we describe the proposed Bayesian game with incomplete information.

Typically, a Bayesian game consists of five parts: i) the set of \emph{players} involved in the game; ii) the set of \emph{players' types}; iii) the set of \emph{players' possible actions}; iv) the set of \emph{subjective probability distributions} representing a player's belief about the other players' types upon knowing his/her type; v) the set of \emph{players' utility functions} providing the payoff of each player given the player's type and a chosen action in the action set \cite{osborne1994course,myerson2013game}.

It is clear that the players involved in the ransomware game are the hacker and the victim. Denote the player set by $\mathcal{N}=\{h,v\}$, where $h$ represents the hacker, and $v$ represents the victim. In the following, we construct the other four parts, i.e., the set of player types, the action set, the set of subjective probability distributions, and the payoff set.

\paragraph{Player types} In the following, we discuss the victim types and hacker types.
\begin{itemize}
  \item[i)] {\em \underline{Victim type}.} Before the hacker encrypts  the files, the victim may or may not be aware of the value of files. We assume that any file can be assigned with some monetary value.  Since different persons or institutions have files with different monetary values, we treat the monetary value of a selected victim's files as some continuous random variable with support on $[0,\infty)$. The $0$ means that the files are inessential to the owner, while $\infty$ represents the files are indispensable and not replaceable. Although in real life, not every person or institution has a clear understanding on the monetary value of the owned data files, different data would  have different values from their own viewpoints. Moreover, once the files are locked by the hacker and  ransom payment is demanded, the victim typically reassesses
the monetary value of the files. Note that how the owner assesses the files should be consistent, no matter whether the files have been locked or not. The hacker is usually unable to know the victim's exact valuation on the data files. Therefore, we assume that the victim has some private information on the monetary value of the data files he/she owns. This private information is referred as the \emph{type  of the victim}, and we denote the type set of the victim by $\mathcal{T}_v=[0,\infty)$.
  \item[ii)] {\em \underline{Hacker type}.}  Whether the ransomware should be paid? There are several factors, among which the following three are typically involved in the decision: i) the value of the encrypted files; ii) the difficulty of recovering the encrypted files (either from the  backup or by cracking the files) without paying the ransom;  iii) the probability of receiving correct information (usually a private key) from the hacker to successfully decrypt the encrypted files. Among those three factors, the iii) is closely related to the hacker's characterization, which can be classified into two groups. The first group is the most common one called the \emph{genuine ransomware hacker}, denoted by $A_1$, whose primary goal is to harvest the victim's ransom. The other group is called the \emph{fake ransomware hacker}, denoted by $A_2$, whose primary goal is to perform the cybercrimes including money laundry, stealing money, and cyber attacks. This group generally does not care much about benefiting from the ransom, and will not response even the victim has paid the ransom.

The reason for us to classify the hackers into the aforementioned groups is supported by the recent survey of \cite{hampton2015ransomware}, where the historical development of ransomware is reviewed and the traits of several major ransomware families are identified. In the survey, it is found that most ransomware families aim at getting the ransom but there are some exceptions. For example, there is a ransomware suite called {\em Reveton}. All variants of this ransomware family tend to use the victim's computer to perform criminal operations, including Bitcoin mining, stealing of Bitcoin wallets, etc. For this ransomware family, demanding a ransom from the victim is not the primary goal. Moreover, \cite{hampton2015ransomware} performed a close examination on the Reveton's features and found that this ransomware family intends to lock the booting process of the victim's computer, instead of directly encrypting the files.

To summary,   we refer those two different groups as the \emph{types} of the hackers. Denote the hacker type set by $\mathcal{T}_h=\{A_1,A_2\}$, where $A_1$ represents the genuine ransomware hacker while $A_2$ represents the fake ransomware hacker.
\end{itemize}

\paragraph{Player actions} In the following, we discuss the possible actions that the hacker and victim may have in the game.
\begin{itemize}
  \item[i)] {\em \underline{Hacker action}.}  When planning the attack, the hacker needs to determine the ransom amount, which is denoted by $r\in\mathcal{A}_h=(0,\infty)$ representing the hacker's action in the game.  We assume that,
after receiving the ransom, the genuine ransomware hacker will send the decryption information while the fake ransomware hacker will not.
  \item[ii)] {\em \underline{Victim  action}.}
Practically, a victim may have several actions. The first one is to discard the files and not pay the ransom (denoted by $D$); the second one is to directly pay the ransom (denoted by $P$); the third one is to hire some technicians to crack the encrypted files and when the cracking fails, to pay the ransom with some probability (denoted by $C$); and the fourth one is to recover the files from backup when this option is available, and when there is no backup or the recovery fails, {to use the cracking strategy $C$} (denoted by $R$). In the sequel,  the action set of the victim is denoted as $\mathcal{A}_v\subset\{D,P,C,R\}$. Note that in practice, the hacker typically adds a ticking countdown timer with some deadline to pay the ransom, and being unable to pay in time will cause the decryption key to be destroyed or the ransom to be increased. Note that it is not an easy task to crack the encrypted files, which may take too much time. Therefore, in this study, we assume that if the victim chooses to crack the encrypted files, then upon its failure, the victim can decide whether to pay the ransom with some extra fee. Meanwhile, cracking and recovery from the backup could be taken sequentially. For instance, a victim having the backup could first choose to recover the files, and when this action fails, the victim will then choose to crack the files.
\end{itemize}

\paragraph{Subjective probabilities.}  Both hacker and victim have the subjective probabilities in the game, which are presented as follows.
\begin{itemize}
  \item[i)] {\em \underline{Victim subjective probability}.}  Once the files are encrypted, the victim is faced with a dilemma (i.e., paying the ransom or not) as the files may not be recovered even the ransom is paid if the hacker is the fake type.  In the real life, even though different types of hackers may prefer different targets, victims are unable to distinguish them since victims  usually do not have expertises in cyber security and social science. Hence, in this study, we assume that the victim's subjective probability on the hacker's type does not depend on how he/she values the files.
   In sum, the subjective probability distribution ${F_v}$ on the hacker's type from the perspective of the victim is a discrete distribution on $\{A_1,A_2\}$ with probability mass $\{p, 1-p\}$.

  \item[ii)] {\em \underline{Hacker subjective probability}.} It would be great for the hacker if he/she knows how the victim values the files. However, due to the fact that different person value files differently, and a person values different types of files differently, the value of files should be treated as a realization of a random variable $V$ following some non-negative distribution. Naturally, the hacker rarely knows this realization.   Hence, we assume that the hacker has some estimation on the value of files, and uses the distribution ${F_h}$  as the valuation distribution. To simplify the discussion, although the hacker has two types,  we assume either type of hacker has the same subjective estimation ${F_h}$ on the  valuation distribution. This simplification is very general  because in practice  both types of hackers make use of system  vulnerabilities to launch the attack, and system vulnerabilities exist across different kinds of victims, regardless how the victim values the files and what the hacker type is.

\end{itemize}

\paragraph{Player utilities}
For each profile of types and actions in $\mathcal{T}_v\times\mathcal{T}_h\times\mathcal{A}_v\times \mathcal{A}_h$, both the hacker and the victim will finally get a ``payoff''. It should be remarked that no matter what profile the victim faces, the victim is doomed to suffer some loss. For simplicity, we still use the word ``payoff'' when referring to the monetary result of a profile of the victim. Generally, the victim's payoff is a function
\[
U_v:\mathcal{T}_v\times\mathcal{T}_h\times\mathcal{A}_v\times \mathcal{A}_h\mapsto (-\infty, 0],
\]
and the larger the payoff is, the better the victim is. Similarly, the hacker's payoff is a function
\[
U_h:\mathcal{T}_v\times\mathcal{T}_h\times \mathcal{A}_v\times \mathcal{A}_h\mapsto (-\infty,\infty).
\]
Note that the hacker's payoff is allowed to be negative. With the negative payoff, the game model takes into account the hacker's total cost (e.g., the resources cost, and the potential risk of being arrested) of planning and launching the ransomware.

\paragraph{Bayesian game}
Now we are ready to introduce the desired Bayesian game to describe the conflicting situation between the hacker and the victim. Although the victim usually does not know the existence of ransomware in his/her computer until the hacker starts the encryption process, the game may actually have started long before. In the present framework, we assume that the game between the hacker and the victim begins at the time that the hacker is planning an attack and selecting some target. The game proceeds at two stages: in the first stage, the hacker decides a ransom and launches the attack; in the second stage, the victim decides whether or not to pay the ransom. Formally, the strategic-form of the concerned game is given by
\begin{equation}\label{game}
\Gamma=\Big(\mathcal{N},(\mathcal{A}_i)_{i\in \mathcal{N}},(\mathcal{T}_i)_{i\in \mathcal{N}},(F_i)_{i\in \mathcal{N}},(U_i)_{i\in \mathcal{N}}\Big).
\end{equation}
In the sequel study of this Bayesian game, we assume that in the entire game, each player knows his/her own type, and  this fact is a  common knowledge between the hacker and the victim. It should be mentioned that although the victim is unaware of the existence of hacker until the files are encrypted, we  allow the victim having the capability of knowing that he/she is involved in the game $\Gamma$.

For the ease of reference, we list in Table \ref{tab01} the mathematical notions vital to the model mentioned in the above discussion.

\begin{table}[!htbp]
\centering
\begin{tabular}{l|l}
\hline
$\mathcal{N}$ & the player set\\ \hline
$\mathcal{T}_v$ & the victim's type set\\ \hline
$\mathcal{T}_h$ & the hacker's type set\\ \hline
$\mathcal{A}_v$ & the victim's action set\\ \hline
$\mathcal{A}_h$ & the hacker's action set\\ \hline
$F_v$ & victim's subjective distribution on the hacker's type\\ \hline
$F_h$ & hacker's subjective distribution on the victim's type\\ \hline
$V$ & the victim's valuation on the encrypted files\\ \hline
$U_v$ & victim's utility function\\ \hline
$U_h$ & hacker's utility function\\ \hline
$D$ & discard the files and not pay the ransom\\ \hline
$P$ & pay the ransom directly\\ \hline
$C$ & crack first, when fail, use $P$ with some probability\\ \hline
$R$ & recover first, when fail, use $C$\\
\hline
\end{tabular}
\caption{Summary of main notations}\label{tab01}
\end{table}

\section{Equilibrium} \label{sec:equi}
In the game theory, Nash equilibrium provides a stable solution to the best strategy problem. Generally, a Nash equilibrium to a game is a tuple of each player's strategy such that no player can be better off by deviating from his/her strategy in this tuple, as long as the other players employ their strategies given in the tuple. In the following, we consider pure and randomized strategies of the ransomware game.

A {\em pure strategy} means that a player with a given type chooses to use the corresponding action determined by the strategy function with probability one. Note that the victim's strategy not only depends on the victim's type, but also on the hacker's strategy. This is because when the victim plans to act, the hacker's strategy is already observable to the victim. For the concerned two-player Bayesian game, denote the pure strategy profile as the tuple $$\big(S_h(\cdot),S_v(\cdot\mid S_h(\cdot))\big),$$ where the function $S_h(\cdot):\mathcal{T}_h\mapsto \mathcal{A}_h$ is the hacker's pure strategy, and the function $S_v(\cdot\mid S_h(\cdot)):\mathcal{T}_v\mapsto\mathcal{A}_v$ is the victim's pure strategy given the hacker's strategy.

A {\em randomized strategy} for any player in a game is a probability distribution over the action set of this player. For the strategic-form game in \eqref{game}, denote $\Delta(\mathcal{A}_i)$ as the set of all possible randomized strategies for the player $i\in\{h,v\}$. { For the victim with four potential actions $D$, $P$, $C$ and $R$, $\Delta(\mathcal{A}_v)$ is a discrete distribution given as
\begin{eqnarray*}
\Delta(\mathcal{A}_v)&=&\big\{(q_D,q_P,q_C,q_R):q_D,q_P,q_C,q_R\in[0,1],\\
&& q_D+q_P+q_C+q_R=1,\\
& &\quad \mbox{$q_i$ is the probability of choosing action $i$ for $i=D,P,C,R$}\big\}.
\end{eqnarray*}
}
For the hacker, $\Delta(\mathcal{A}_h)$ consists of any continuous distribution supported on $(0,\infty)$ given as
\[
\Delta(\mathcal{A}_h)=\big\{G:\mbox{$G$ is a continuous distribution function supported on $(0,\infty)$}\}.
\]

In a Bayesian game with incomplete information, the \emph{Bayesian Nash equilibrium}  defined in  \cite{harsanyi1967games,harsanyi1968games} is any Nash equilibrium (of pure strategy, or randomized strategy) specifying a strategy which maximizes the corresponding expected payoff\footnote{The expectation is taken under the subjective probability measure of each player.} for each type of player, given that the concerned player only knows his/her own type but not those of the other players. Below we formally introduce the definition of a randomized-strategy Bayesian Nash equilibrium.

\begin{Def}\rm
In an $n$-player Bayesian game
\[
\Gamma^b=\Big(\mathcal{N},(\mathcal{A}_i)_{i\in \mathcal{N}},(\mathcal{T}_i)_{i\in \mathcal{N}},(\mathcal{F}_i)_{i\in \mathcal{N}},(U_i)_{i\in \mathcal{N}}\Big),
\]
consider a strategy profile
\[
\bm q=\bigg(\Big(q_i(x\mid t)\Big)_{x\in\mathcal{A}_i}\bigg)_{t\in\mathcal{T}_i,i\in\mathcal{N}}
\]
defined as a probability distribution over the product space
\begin{equation}\label{eq05}
\mbox{\huge{$\times$}}_{i\in\mathcal{N}}\mbox{\huge{$\times$}}_{t\in\mathcal{T}_i}
\Delta_{t}(\mathcal{A}_i),
\end{equation}
where $\Delta_{t}(\mathcal{A}_i)$ is the set of all possible randomized strategies for the player $i$ given this player knowing his/her own type $t\in\mathcal{T}_i$, and $\big(q_i(x\mid t)\big)_{x\in\mathcal{A}_i}$ is a conditional probability distribution over $\mathcal{A}_i$ given the player $i$ knowing his/her own type $t\in\mathcal{T}_i$. Then, $\bm q$ is a \emph{randomized-strategy} Bayesian Nash equilibrium if, for each player $i\in\mathcal{N}$ and each type $t\in\mathcal{T}_i$,
\begin{equation}\label{bayeq}
\Big(q_i(x\mid t)\Big)_{x\in\mathcal{A}_i}\in \mathop{\arg\max}\limits_{\tau\in\Delta(\mathcal{A}_i)}\mathsf{E}\bigg[\sum_{\bm a\in(\mathcal{A}_i)_{i\in \mathcal{N}}}\prod_{j\in\mathcal{N}\backslash\{i\}}q_{j}(a_j\mid \mathcal{T}_j)\tau(a_i)U_i(\mathcal{T}_1,\cdots,\mathcal{T}_n,\bm a)\bigg\vert \mathcal{T}_i=t\bigg].
\end{equation}
\end{Def}

Note that in our setting, a randomized Bayesian Nash equilibrium of the game $\Gamma$ in \eqref{game} is a profile $\bm q$ such that, for each player $i\in\mathcal{N}$ and each type $t\in\mathcal{T}_i$,
\begin{equation}\label{bayeq}
\Big(q_i(x\mid t)\Big)_{x\in\mathcal{A}_i}\in \mathop{\arg\max}\limits_{\tau\in\Delta(\mathcal{A}_i)}\mathsf{E}\bigg[
\sum_{(x,y)\in\mathcal{A}_i\times\mathcal{A}_j}q_{j}(y\mid \mathcal{T}_j)\tau(x)U_i(t,\mathcal{T}_j,x,y)\bigg].
\end{equation}
where $\{i,j\}=\mathcal{N}=\{h,v\}$.

To derive the Bayesian Nash equilibrium explicitly, we need to specify exogenous parameters of the proposed framework.

For the victim, assume that the probability of successful cracking the files without the decryption information is $p_1$, and successful or not, cracking files costs the victim $c_1$, regardless of the type of the hacker. Moreover, if the victim fails to crack the files, then
the victim pays the ransom with an extra punishment fee $c_2\ge 0$ with probability $p_2$. To circumstance potential complexity, we assume that this probability $p_2$ depends on only one endogenous parameter $r$ and some other possible exogenous parameters known to the hacker and the victim. Further,  it is reasonable to assume that this probability is continuously non-increasing 
 in the ransom $r$ and satisfies some basic requirements, i.e.,
\[
p_2:(0,\infty)\mapsto [0,1]
\]
such that $\lim\limits_{r\to0^+}p_2(r)\le 1$ and $\lim\limits_{r\to\infty}p_2(r)=0$. In this study, we assume the punishment fee is not paid to the hacker and is just some monetary measurement on the victim's intension level due to the elapsed time of cracking. If the victim decides to recover the files from the backup, there exists a probability $p_3$ that the recovery is successful, and  it costs $c_3$ for the recovery no matter the recovery is successful or not.

For the hacker, it costs a hacker (both types)  amount $c_4$ to plan and launch the attack. By manipulating the victim's computer, a type $A_2$ hacker is able to earn $b_1$ even if the victim discards the files, the encrypted files are successfully cracked by the victim {or the files are successfully recovered,} and earn $b_2$ (excluding the ransom) if the victim chooses to pay the ransom, fails to crack the files {or fails to recover the files.} Since the victim's computer can be used for Bitcoin mining, or the hacker is stealing password and credential or Bitcoin wallets, the longer the computer is under control of the hacker, the more the hacker will benefit   \cite{Kujawa2013,hampton2015ransomware,Saarinen2015}.   Hence, it is reasonable to assume $b_1\le b_2$, and there is no cost  for the hacker to send back the decryption message. Table \ref{tab02} summarizes the aforementioned notations.
\begin{table}[!htbp]
\centering
\begin{tabular}{l|l}
\hline
$p$ & the probability of encountering type $A_1$ hacker \\ \hline
$p_1$ & the probability of successfully cracking the files without the decryption information\\ \hline
$p_2$ & the probability that the victim pays the ransom with an extra  punishment fee $c_2$,\\
& which is a continuously non-increasing function in $r$ with $\lim\limits_{r\to0^+}p_2(r)\le 1$ and $\lim\limits_{r\to\infty}p_2(r)=0$\\ \hline
$p_3$ &  the probability of successfully recovering from the backup\\ \hline
$c_1$ &  the cost of cracking files\\ \hline
$c_2$ &   the extra  cost  upon the failure of cracking files\\ \hline
$c_3$ &  the cost of recovering files from the backup\\ \hline
$c_4$ &  the cost of planning and launching the ransomware attack\\ \hline
$b_1$ & the earning of type $A_2$ hacker if the victim discards the files, \\
& {the encrypted files are successfully cracked or recovered by the victim}\\\hline
$b_2$ & the earning of type $A_2$ hacker (excluding the ransom) if the victim chooses to pay the ransom,\\
& {fails to crack or recover the encrypted files}\\ \hline
\end{tabular}
\caption{Parameters in the game}\label{tab02}
\end{table}

In practice, the encryption technology used by the hacker is normally extremely powerful, and without a proper private key, the probability $p_1$  of decrypting an encrypted file  is very small. Since most hackers will send back the correct private key after receiving the ransom in the reported ransomware attacks \cite{hampton2015ransomware}, it is assumed that $p>p_1$.

We assume that all the probabilities $(p,p_1,p_2,p_3)$ and costs $(c_1,c_2,c_3)$ are common knowledge between the hacker and the victim, whenever they are included in the model framework. In practice, the proportion $p$ of type $A_1$ hacker can be estimated from some reviewing reports on ransomware families  \cite{conti2018economic}.  The probability $p_2$ measures the victim's willingness to pay the ransom after  the failure of cracking files. The successful recovery probability $p_3$ may be estimated by the proportion of institutions using online storage found in public reports; $c_1,c_3$ could be the market price of similar data decryption business and data recovery business accessible to both the victim and the hacker, respectively;  $c_2$ could be taken as some fraction of the cracking cost. Note that the cost $c_4$ of launching an attack is not assumed to be a common knowledge due to the fact that a victim usually lacks of information on hackers.

As for type $A_2$ hacker's earning ability ($b_1$ and $b_2$), we assume that the victim is unable to obtain this information, and only knows their existence but not the exact values. At the first glance, this information may have some effect on the victim's decision. In the later discussion, we will show that it does not play a role in the victim's best response, partially because that to make a decision, a victim just needs to know the ransom amount and the valuation on the files.

In the following sections, we study the existence of Bayesian Nash equilibrium of the proposed game, and explore the effects of different parameters in the game on the equilibrium. We assume that the hacker will not plan and launch a ransomware attack to some potential victim if he/she has a  non-positive expected payoff of the attack. 

\subsection{Scenario I: victim has no backup}\label{case1}
We first discuss the scenario that the victim does not have the backup for the files. In this case, we assume that after the files are encrypted, the victim can choose to discard the files, pay the ransom directly, or try to crack the files and pay the ransom when the cracking action fails\footnote{The encryption technology used by the hacker has been developed over several decades, and using brute force to decrypt the encrypted files usually consumes an extremely large computation resource, to which most victims can hardly access. The decryption information is usually some private keys and it is possible to capture this information when the hacker transmits it to the victim. With the combined effort of private and government organizations, a large number of private keys has been collected, and when the files are encrypted, a victim can try all of the available ones to see whether there is a match. Although this makes it more possible to decrypt the files without paying the ransom, there is still no guarantee.}. That is, the victim's action set is given by $\mathcal{A}_v=\{D,P,C\}$, where $D$ denotes that the victim discards the files, $P$ denotes that the victim pays the ransom, and $C$ denotes that the victim chooses to crack the files first and only when this action fails, to pay the ransom with some probability. For the ease of reference, denote $\Gamma_1$ as the strategic-form of the game \eqref{game} under this case.

The following theorem reveals when the victim prefers to pay the ransom. Note that when the expected payoffs of using different strategies are the same, we assume that the victim always prefers not to pay the ransom {and discard the files}, i.e., $D$; when the expected payoffs of $P$ and $C$ are the same and greater than that of $D$, we assume that the victim always prefers $C$,  to avoid tedious mathematical derivation.

Before presenting the theorem, we first introduce the following lemma.

\begin{Lem}\rm\label{lem00}
Denote
\begin{equation}\label{eq:psi1}
\psi_1(r)=\frac{c_1+(c_2+r)p_2(r)(1-p_1)-r}{(1-p)[1-p_2(r)(1-p_1)]},
\end{equation}
and
\begin{equation}\label{eq:psi2}
\psi_2(r)=\frac{c_1+(c_2+r)p_2(r)(1-p_1)}{1-p_2(r)(1-p_1)(1-p)}.
\end{equation}
Let
\begin{eqnarray*}
& &\mathcal{M}_1=\big\{r\ge 0:r/p\le\psi_2(r)\le \psi_1(r)\big\},\quad\mathcal{M}_2=\big\{r\ge 0:\psi_1(r)\le\psi_2(r)\le r/p\big\}.
\end{eqnarray*}
Then $\mathcal{M}_1=[0,\omega_0]$ and $\mathcal{M}_2=[\omega_0,\infty)$, {where $\omega_0$ is the solution to the equation
\[
c_1p+c_2p_2(r)(1-p_1)p+rp_2(r)(1-p_1)-r=0.
\]}
\end{Lem}

{\bf Proof:} {Since the probability function of  $p_2(r)$ is continuous in $r$, it holds that $\psi_1(r)$ and $\psi_2(r)$ are both continuous in $[0,+\infty)$. Moreover,
\begin{eqnarray*}
\psi_1(r)-\frac{r}{p}&\stackrel{\text{sgn}}{=}&\psi_2(r)-\frac{r}{p}\\
&\stackrel{\text{sgn}}{=}&\psi_1(r)-\psi_2(r)\\
&
\stackrel{\text{sgn}}{=}&c_1p+c_2p_2(r)(1-p_1)p+rp_2(r)(1-p_1)-r\\
&=&\Psi(r),
\end{eqnarray*}
where $\stackrel{\text{sgn}}{=}$ represents both sides have the same sign. Note that the non-increasing property of $p_2(r)$ implies that $\Psi(r)$ is also decreasing in $r$. Then, we have
\[
\Psi(0)=c_1p+c_2p_2(0)(1-p_1)p\ge \Psi(r)\ge \lim_{r\to\infty}\Psi(r)=-\infty.
\]
Therefore, the equation $\Psi(r)=0$ must have at least one solution, denoted as $\omega_0$, such that $\Psi(r)\ge 0$ for $r\in[0,\omega_0]$, and $\Psi(r)\le 0$ for $r\in[\omega_0,\infty)$. Therefore, $\mathcal{M}_1=[0,\omega_0]$ and $\mathcal{M}_2=[\omega_0,\  infty)$. This completes the proof.} \qed

{\bf Remark:}  $\mathcal{M}_1$ represents the set of small ransom amount while $\mathcal{M}_2$ represents the set of large ransom amount, and $\omega_0$ is the threshold value between two sets.

\medskip
Based on Lemma \ref{lem00}, we present the victim's weakly dominant strategy when facing a ransom demand from the hacker in the following theorem.

\begin{The}\rm\label{the01}
For the Bayesian game $\Gamma_1$, given the ransom demand $r$,
\begin{enumerate}[(i)]
\item if $r\in\mathcal{M}_1$, then it is weakly dominant for the victim to use the strategy
\begin{equation}\label{eq03}
S_v(x)=\left\{
         \begin{array}{ll}
           D, & x\le r/p, \\
           P, & r/p<x\le\psi_1(r),\\
           C, & x>\psi_1(r);
         \end{array}
       \right.
\end{equation}
\item if $r\in\mathcal{M}_2$, then it is weakly dominant for the victim to use the strategy
\begin{equation*}
S_v(x)=\left\{
         \begin{array}{ll}
           D, & x\le \psi_2(r), \\
           C, & x>\psi_2(r),
         \end{array}
       \right.
\end{equation*}
and the victim should never pay the ransom first.
\end{enumerate}
\end{The}

{\bf Proof:} Note that, for a victim with private type $x\in[0,\infty)$ (i.e., the valuation on encrypted files), the utility functions are given by:
\begin{eqnarray*}
& &U_v(x,A_1,D,r)=U_v(x,A_2,D,r)=-x,\\
& &U_v(x,A_1,P,r)=-r,\quad U_v(x,A_2,P,r)=-x-r,\\
& &U_v(x,A_1,C,r)=-c_1-(c_2+r)p_2(r)(1-p_1),\\
& &U_v(x,A_2,C,r)=-c_1-(x+c_2+r)p_2(r)(1-p_1).
\end{eqnarray*}

Let $T_h$ be the type of the hacker,  then
$$\mathrm{P}(T_h=A_1)=p,\quad\mathrm{P}(T_h=A_2)=1-p.$$
If the strategy $S_v(x)=D$ is used, then
\[
\mathsf{E}[U_v(x,T_h,D,r)]=-x.
\]
If the strategy $S_v(x)=P$ is used, then
\[
\mathsf{E}[U_v(x,T_h,P,r)]=U_v(x,A_1,P,r)p+U_v(x,A_2,P,r)(1-p)=-x(1-p)-r.
\]
If the strategy $S_v(x)=C$ is used, then
\begin{eqnarray*}
& &\mathsf{E}[U_v(x,T_h,C,r)]\\
&=&U_v(x,A_1,C,r)p+U_v(x,A_2,C,r)(1-p)\\
&=&-xp_2(r)(1-p_1)(1-p)-c_1-(c_2+r)p_2(r)(1-p_1).
\end{eqnarray*}
Therefore, the difference between expected payoffs using pure strategies $D$ and $P$ is
\begin{eqnarray}\label{eq01a}
& &\mathsf{E}[U_v(x,T_h,D,r)]-\mathsf{E}[U_v(x,T_h,P,r)]\notag\\
&=&-x-[-x(1-p)-r]\notag\\
&\stackrel{\text{sgn}}{=}&-x+r/p.
\end{eqnarray}
The difference between expected payoffs using pure strategies $P$ and $C$ is
\begin{eqnarray}\label{eq01b}
& &\mathsf{E}[U_v(x,T_h,P,r)]-\mathsf{E}[U_v(x,T_h,C,r)]\notag\\
&=&[-x(1-p)-r]-[-xp_2(r)(1-p_1)(1-p)-c_1-(c_2+r)p_2(r)(1-p_1)]\notag\\
&=&-x(1-p)[1-p_2(r)(1-p_1)]+c_1+(c_2+r)p_2(r)(1-p_1)-r\notag\\
&\stackrel{\text{sgn}}{=}&-x+\psi_1(r).
\end{eqnarray}
Similarly, the difference between expected payoffs using pure strategies $D$ and $C$ is
\begin{eqnarray}\label{eq01c}
& &\mathsf{E}[U_v(x,T_h,D,r)]-\mathsf{E}[U_v(x,T_h,C,r)]\notag\\
&=&-x-[-xp_2(r)(1-p_1)(1-p)-c_1-(c_2+r)p_2(r)(1-p_1)]\notag\\
&=&-x[1-p_2(r)(1-p_1)(1-p)]+c_1+(c_2+r)p_2(r)(1-p_1)\notag\\
&\stackrel{\text{sgn}}{=}&-x+\psi_2(r).
\end{eqnarray}

We only prove the case $r\in\mathcal{M}_1$, and the case of (ii) can be verified similarly. Note that the victim always prefers a strategy with a larger expected payoff. For $r\in\mathcal{M}_1$,  according to \eqref{eq01a} and \eqref{eq01c}, it holds that, for $x\le r/p$,
\[
\mathsf{E}[U_v(x,T_h,D,r)]\ge\mathsf{E}[U_v(x,T_h,P,r)],\quad\mathsf{E}[U_v(x,T_h,D,r)]\ge\mathsf{E}[U_v(x,T_h,C,r)].
\]
When $r/p<x\le\psi_1(r)$, it follows from \eqref{eq01a} and \eqref{eq01b} that
\[
\mathsf{E}[U_v(x,T_h,P,r)]\ge\mathsf{E}[U_v(x,T_h,D,r)],\quad\mathsf{E}[U_v(x,T_h,P,r)]\ge\mathsf{E}[U_v(x,T_h,C,r)].
\]
If $x>\psi_1(r)$, then it follows from \eqref{eq01b} and $\eqref{eq01c}$ that
\[
\mathsf{E}[U_v(x,T_h,C,r)]\ge\mathsf{E}[U_v(x,T_h,P,r)],\quad\mathsf{E}[U_v(x,T_h,C,r)]\ge\mathsf{E}[U_v(x,T_h,D,r)].
\]

That is, the strategy in Eq. \eqref{eq03} leads to a higher expected payoff for the victim. \qed

{\bf Remark:} Theorem \ref{the01} shows that when the valuation on encrypted files is much less than the ransom amount, the victim should choose to discard the files directly; when the valuation is large compared to the ransom amount, the victim should try to crack the files first, which will result in a positive probability of not paying the ransom; {\em when the ransom amount is not too large and the victim's valuation is neither too small nor too large, the victim should choose to pay the ransom.}
Based on Lemma \ref{lem00} and Theorem \ref{the01}, for a victim with valuation $x$, the weakly dominant strategy can be concisely expressed as
\[
D\mathbb{I}(x\le r/p\wedge \psi_2(r))+C\mathbb{I}(x>\psi_1(r)\vee \psi_2(r))+P\mathbb{I}(r/p<x\le \psi_1(r)),
\]
{where $`\wedge'$ represents the minimum, and `$\vee$' represents the maximum.}

Note the hacker's utility is affected by whether the victim pays the ransom. For the hacker with private type $A_1$, the utility functions are
\[
U_h(x,A_1,D,r)=-c_4,\quad U_h(x,A_1,P,r)=r-c_4,\quad U_h(x,A_1,C,r)=rp_2(r)(1-p_1)-c_4.
\]
For the hacker with private type $A_2$, the utility functions are as follows,
\begin{eqnarray*}
& &U_h(x,A_2,D,r)=b_1-c_4,\quad U_h(x,A_2,P,r)=b_2+r-c_4,\\
& &U_h(x,A_2,C,r)=b_1p_1+(b_2+rp_2(r))(1-p_1)-c_4.
\end{eqnarray*}
In the next theorem, we show that {\em there exists a pure Bayesian Nash equilibrium for the hacker and the victim in the Bayesian game $\Gamma_1$}.  To facilitate the discussion of equilibrium, we allow the ransom amount to be $0$, which means that the hacker will not launch the ransomware attack.  Before presenting the result on the Bayesian Nash equilibrium, we first present the following lemma.

\begin{Lem}\rm\label{lem01} Let $r_1^*$ be the smallest maximizer to
\begin{eqnarray*}
\eta_1(r)&=&rp_2(r)(1-p_1)\big[\bar F_h(\psi_1(r))\mathbb{I}\big(r\in\mathcal{M}_1\big)+\bar F_h(\psi_2(r))\mathbb{I}\big(r\in\mathcal{M}_2\big)\big]\\
& &+r\big[\bar F_h(r/p)-\bar F_h(\psi_1(r))\big]\mathbb{I}\big(r\in\mathcal{M}_1\big),
\end{eqnarray*}
and $r_2^*$ be the smallest maximizer to
\begin{eqnarray*}
\eta_2(r)&=&[b_2+(b_1-b_2)p_1+rp_2(r)(1-p_1)]\big[\bar F_h(\psi_1(r))\mathbb{I}\big(r\in\mathcal{M}_1\big)+\bar F_h(\psi_2(r))\mathbb{I}\big(r\in\mathcal{M}_2\big)\big]\\
& &+(b_2+r)\big[\bar F_h(r/p)-\bar F_h(\psi_1(r))\big]\mathbb{I}\big(r\in\mathcal{M}_1\big)\\
& &+b_1\big[F_h(r/p)\mathbb{I}\big(r\in\mathcal{M}_1\big)+F_h(\psi_2(r))\mathbb{I}\big(r\in\mathcal{M}_2\big)\big],
\end{eqnarray*}
where $\bar F_h(\cdot)=1-F_h(\cdot)$, i.e., the survival function of hacker's subjective valuation on the victim's files.
If the probability $p_2(r)$ satisfies the following finite condition \begin{equation}\label{con1}
\lim_{r\to\infty}rp_2(r)\le \ell,
\end{equation}
where $\ell \ge 0$ is some constant, then $r_1^*$ and $r_2^*$ must exist.
\end{Lem}

{\bf Proof:} We only verify the existence of $r_1^*$, and that of $r_2^*$ can be proved similarly.  Note that, for  $r\in\mathcal{M}_1$, it holds that
 $$\eta_1(r)=rp_2(r)(1-p_1)\bar F_h(\psi_1(r))+ r\big[\bar F_h(r/p)-\bar F_h(\psi_1(r))\big].$$
 Since any continuous function has a local maximum on a given closed interval,  $\eta_1(r)$ has a local maximizer $\tilde r_1$ on $[0,\omega_0]$.
For  $r\in\mathcal{M}_2$, we have
\begin{equation}\label{eq02a}
\eta_1(r)=rp_2(r)(1-p_1)\bar F_h(\psi_2(r)).
\end{equation}
Moreover, condition \eqref{con1} guarantees that there exists some  $\hat\omega_0>0$ such that for $r\ge \hat\omega_0$, $rp_2(r)\le \ell$, implying that \eqref{eq02a} has some local maximizer $\hat r_1$ on $[ \hat\omega_0,\infty)$. Because of the  continuity of  $\eta_1(r)$ on $ [\omega_0, \hat\omega_0]$,  there also exists a local maximizer.  Therefore, the conclusion follows immediately.
\qed

\begin{The}\rm\label{the02}
Assume that  $p_2(\cdot)$ fulfills condition \eqref{con1}. For the Bayesian game $\Gamma_1$, there exists a pure Bayesian Nash equilibrium such that the hacker and the victim use the strategy profile $\Big(S_h^e(A_i),S_v^e\big(\cdot\mid S_h^e(A_i)\big)\Big)$, where
\[
S_h^e(A_1)=r_1^*\mathbb{I}\big(\eta_1(r_1^*)>c_4\big),\quad\mbox{and}\quad S_h^e(A_2)=r_2^*\mathbb{I}\big(\eta_2(r_2^*)>c_4\big),
\]
and for any $x\in[0,\infty)$,
\begin{eqnarray}\label{eq09}
& &S_v^e\big(x\mid S_h^e(\cdot)\big)\\
&=&D\Big[\mathbb{I}\Big(x\le S_h^e(\cdot)/p\wedge\psi_2\big(S_h^e(\cdot)\big)\Big)\Big]+P\mathbb{I}\Big(S_h^e(\cdot)/p\le x\le \psi_1\big(S_h^e(\cdot)\big)\Big)\notag\\
& &+C\Big[\mathbb{I}\Big(x>\psi_1\big(S_h^e(\cdot)\big)\vee\psi_2\big(S_h^e(\cdot)\big)\Big)\Big].\notag
\end{eqnarray}
\end{The}

{\bf Proof:} Theorem \ref{the01} shows that $S_v^e\big(x\mid r\big)$ is a weakly dominant strategy of a victim with type $x$ given any strategy $r$ used by the hacker. Hence, it is enough to consider the best strategy of the hacker under the scenario that the victim uses the strategy $S_v^e\big(x\mid r\big)$.  Now, for $S_h(A_1)=r$, we have
\begin{eqnarray}\label{eq02}
& &\mathsf{E}\big[U_h\big(V,A_1,S_v(V),r\big)\big]\notag\\
&=&\sum_{i=1}^2\mathsf{E}\big[U_h\big(V,A_1,S_v(V),r\big)\big]\mathbb{I}\big(r\in\mathcal{M}_i\big)\notag\\
&=&\mathbb{I}\big(r\in\mathcal{M}_1\big)\bigg[\int_0^{r/p}U_h\big(x,A_1,D,r\big)\dif F_h(x)+\int_{r/p}^{\psi_1(r)}U_h\big(x,A_1,P,r\big)\dif F_h(x)\notag\\
& &\qquad+\int_{\psi_1(r)}^{\infty}U_h\big(x,A_1,C,r\big)\dif F_h(x)\bigg]\notag\\
& &+\mathbb{I}\big(r\in\mathcal{M}_2\big)\bigg[\int_0^{\psi_2(r)}U_h\big(x,A_1,D,r\big)\dif F_h(x)+\int_{\psi_2(r)}^{\infty}U_h\big(x,A_1,C,r\big)\dif F_h(x)\bigg]\notag\\
&=&rp_2(r)(1-p_1)\big[\bar F_h(\psi_1(r))\mathbb{I}\big(r\in\mathcal{M}_1\big)+\bar F_h(\psi_2(r))\mathbb{I}\big(r\in\mathcal{M}_2\big)\big]\notag\\
& &+r\big[\bar F_h(r/p)-\bar F_h(\psi_1(r))\big]\mathbb{I}\big(r\in\mathcal{M}_1\big) -c_4\notag\\
&=&\eta_1(r)-c_4.
\end{eqnarray}
Similarly, for $S_h(A_2)=r$, we have
\begin{eqnarray}\label{eq04}
& &\mathsf{E}\big[U_h\big(V,A_2,S_v(V),r\big)\big]\notag\\
&=&\sum_{i=1}^2\mathsf{E}\big[U_h\big(V,A_2,S_v(V),r\big)\big]\mathbb{I}\big(r\in\mathcal{M}_i\big)\notag\\
&=&[b_2+(b_1-b_2)p_1+rp_2(r)(1-p_1)]\big[\bar F_h(\psi_1(r))\mathbb{I}\big(r\in\mathcal{M}_1\big)+\bar F_h(\psi_2(r))\mathbb{I}\big(r\in\mathcal{M}_2\big)\big]\notag\\
& &+(b_2+r)\big[\bar F_h(r/p)-\bar F_h(\psi_1(r))\big]\mathbb{I}\big(r\in\mathcal{M}_1\big)\notag\\
& &+b_1\big[F_h(r/p)\mathbb{I}\big(r\in\mathcal{M}_1\big)+F_h(\psi_2(r))\mathbb{I}\big(r\in\mathcal{M}_2\big)\big]-c_4\notag\\
&=&\eta_2(r)-c_4.
\end{eqnarray}

According to Eq. \eqref{eq02} and Lemma \ref{lem01}, the hacker's maximum expected payoff is $\eta_1(r_1^*)-c_4$. Since the hacker will not launch the attack if the expected payoff is not strictly positive, the best strategy for type $A_1$ hacker must be  $S_h^e(A_1)=r_1^*\mathbb{I}\big(\eta_1(r_1^*)>c_4\big)$. Similarly,
the best strategy for type $A_2$ hacker must be $S_h^e(A_2)=r_2^*\mathbb{I}\big(\eta_2(r_2^*)>c_4\big)$. This completes the proof.
\qed

{\bf Remark:} Theorem \ref{the02} shows that in the equilibrium, the ransom amount demanded by the hacker will not be too small. That is, the hacker should have enough incentive to launch the ransomware attack. If the hacker asks for a `correct' ransom amount, the victim may choose to directly pay the ransom, which depends on the victim's valuation on the encrypted files. Further,  type $A_2$ hacker always attains a larger expected payoff than type $A_1$ hacker does, because for all $r\ge 0$,
\begin{eqnarray*}
\eta_2(r)-\eta_1(r)&=&[b_2(1-p_1)+b_1p_1]\big[\bar F_h(\psi_1(r))\mathbb{I}\big(r\in\mathcal{M}_1\big)+\bar F_h(\psi_2(r))\mathbb{I}\big(r\in\mathcal{M}_2\big)\big]\\
& &+b_2\big[\bar F_h(r/p)-\bar F_h(\psi_1(r))\big]\mathbb{I}\big(r\in\mathcal{M}_1\big)\\
& &+b_1\big[F_h(r/p)\mathbb{I}\big(r\in\mathcal{M}_1\big)+F_h(\psi_2(r))\mathbb{I}\big(r\in\mathcal{M}_2\big)\big]\\
&\ge&0.
\end{eqnarray*}

The following result compares the ransom amounts demanded by type $A_1$ hacker and type $A_2$ hacker in the equilibrium.

\begin{The}\rm\label{the05}
Assume $rp_2(\cdot)$ is decreasing. 
For the Bayesian game $\Gamma_1$ in equilibrium,  let the equilibrium ransom amounts of hacker $A_1$ and $A_2$ be  $r_1^*$ and $r_2^*$, respectively.

\begin{enumerate}[(i)]
\item If $r_2^*\in\mathcal{M}_1$, then $r_1^*\ge r_2^*$.
\item If $r_2^*\in\mathcal{M}_2$, then $r_1^*\le r_2^*$.
\end{enumerate}
\end{The}

{\bf Proof:} We only prove the case (i) and the other case can be proved similarly. According to Eqs. \eqref{eq02} and \eqref{eq04}, given the same ransom amount $r$, the difference between expected payoffs of type $A_2$ hacker and type $A_1$ hacker is
\begin{eqnarray*}\label{eq13}
d(r)&= &\mathsf{E}\big[U_h\big(V,A_2,S_v(V),r\big)\big]-\mathsf{E}\big[U_h\big(V,A_1,S_v(V),r\big)\big]\notag\\
&=&[b_2+(b_1-b_2)p_1]\big[\bar F_h(\psi_1(r))\mathbb{I}\big(r\in\mathcal{M}_1\big)+\bar F_h(\psi_2(r))\mathbb{I}\big(r\in\mathcal{M}_2\big)\big]\notag\\
& &+b_2\big[\bar F_h(r/p)-\bar F_h(\psi_1(r))\big]\mathbb{I}\big(r\in\mathcal{M}_1\big)\notag\\
& &+b_1\big[F_h(r/p)\mathbb{I}\big(r\in\mathcal{M}_1\big)+F_h(\psi_2(r))\mathbb{I}\big(r\in\mathcal{M}_2\big)\big].
\end{eqnarray*}
Note that $\mathcal{M}_1=[0,\omega_0]$, and $\mathcal{M}_1=[\omega_0,\infty)$. It holds that
\begin{eqnarray}\label{eq13b}
d(r)&=&\left\{
   \begin{array}{ll}
     b_2+(b_1-b_2)\big[p_1\bar F_h(\psi_1(r))+F_h(r/p)\big], & r\in[0,\omega_0],\\
     b_1+(1-p_1)(b_2-b_1)\bar F_h(\psi_2(r)), & r\in[\omega_0,\infty).
   \end{array}
 \right.
\end{eqnarray}

Note that the decreasing property of $rp_2(r)$ implies $p_2(r)$ is decreasing, and hence $\psi_1(r)$ and $\psi_2(r)$ are decreasing in $r$.  Further, since $b_1\le b_2$, it follows that the difference $d(r)$ in Eq. \eqref{eq13b} is decreasing in $r\le \omega_0$ and increasing in $r\ge \omega_0$. Consider the equilibrium ransom amount $r_2^*$ of type $A_2$ hacker. If $r_2^*\in\mathcal{M}_1$, i.e., $r_2^*\le \omega_0$, then for $r\le r_2^*$,
\begin{eqnarray*}
& &\mathsf{E}\big[U_h\big(V,A_2,S_v(V),r\big)\big]-\mathsf{E}\big[U_h\big(V,A_1,S_v(V),r\big)\big]\\
&\ge&\mathsf{E}\big[U_h\big(V,A_2,S_v(V),r_2^*\big)\big]-\mathsf{E}\big[U_h\big(V,A_1,S_v(V),r_2^*\big)\big].
\end{eqnarray*}
Therefore, it follows that,  for any $r\le r_2^*\le \omega_0$,
\begin{eqnarray*}
& &\mathsf{E}\big[U_h\big(V,A_1,S_v(V),r\big)\big]\\
&\le&\mathsf{E}\big[U_h\big(V,A_2,S_v(V),r\big)\big]-\mathsf{E}\big[U_h\big(V,A_2,S_v(V),r_2^*\big)\big]+\mathsf{E}\big[U_h\big(V,A_1,S_v(V),r_2^*\big)\big]\\
&\le&\mathsf{E}\big[U_h\big(V,A_1,S_v(V),r_2^*\big)\big].
\end{eqnarray*}
Note that the equilibrium ransom amount $r_1^*$ of type $A_1$ hacker implies that
\[
\mathsf{E}\big[U_h\big(V,A_1,S_v(V),r_1^*\big)\big]\ge \mathsf{E}\big[U_h\big(V,A_1,S_v(V),r_2^*\big)\big].
\]
Therefore,  $r_1^*$ must be greater than  or equal to $r_2^*$. Hence, the desired conclusion follows. \qed

{\bf Remark:} Theorem \ref{the05} implies that in the equilibrium, if the ransom amount demanded by type $A_2$ hacker is relatively small, then the ransom amount demanded by type $A_1$ hacker is larger. However, if the ransom amount demanded by type $A_2$ hacker is relatively large, then the ransom amount demanded by type $A_1$ hacker is smaller. This can be explained by the fact that type $A_2$ hacker is able to earn some profit no matter what action the victim takes. Further, the victim is more willing to pay a small ransom but to crack when faced with a relatively large ransom, which further benefits type $A_2$ hacker. 

\medskip
Theorem \ref{the02} provides one pure Bayesian Nash equilibrium of the game $\Gamma_1$. In fact, using the similar argument, we can prove a randomized Bayesian Nash equilibrium for this game when the maximizers to $\eta_1(r)=\mathsf{E}\big[U_h\big(V,A_1,S_v(V),r\big)\big]+c_4$ and $\eta_2(r)=
\mathsf{E}\big[U_h\big(V,A_2,S_v(V),r\big)\big]+c_4$ are not unique, respectively.

\begin{The}\rm\label{cor01}
Assume $p_2(\cdot)$ fulfills condition \eqref{con1}. For the Bayesian game $\Gamma_1$, if $\mathcal{R}_1=\mathop{\arg\max}_r\eta_1(r)$ and $\mathcal{R}_2=\mathop{\arg\max}_r\eta_2(r)$ are both non-singleton, then the randomized Bayesian Nash equilibrium of the game is given by $\Big(S_h^e(A_i),S_v^e\big(\cdot\mid S_h^e(A_i)\big)\Big)$, with
\[
S_h^e(A_1)=\sigma_1\mathbb{I}\big(\eta_1(r_1^*)>c_4\big),\quad\mbox{and}\quad S_h^e(A_2)=\sigma_2\mathbb{I}\big(\eta_2(r_2^*)>c_4\big)
\]
and $S_v^e\big(x\mid S_h^e(\cdot)\big)$ is given by \eqref{eq09}, where
\[
\bigg\{\sigma_i:\mathcal{R}_i\mapsto [0,1]:\sum_{r\in\mathcal{R}_i}\sigma_i(r)=1\bigg\},
\]
for $i=1,2$.
\end{The}

{\bf Proof:} Note that for type $A_1$ ($A_2$) hacker, any strategy in $\mathcal{R}_1$ ($\mathcal{R}_2$) achieves the same maximum expected payoff $\eta_1(r_1^*)-c_4$ ($\eta_2(r_2^*)-c_4$). Hence, any distribution over the set $\mathcal{R}_i$ results in the same expected payoff for the hacker, $i=1,2$. Moreover, Theorem \ref{the01} shows that the victim's best response to a given ransom claim is always given by Eq. \eqref{eq09}. Therefore, the required result follows immediately. \qed

In the following example, we  illustrate the best response of the victim given any ransom and the pure equilibria of both types of hackers.

\begin{Exa}\rm  \label{exa03} Assume that in game $\Gamma_1$, the parameter setting is $p=0.9$, $p_2(r)=(1+r)^{-2}$, $b_1=1$, $b_2=1.5$, $c_1=1$, $c_2=0.5$ and $c_4=0.2$. We also assume that the hacker's subjective probability distribution on the victim's valuation is the standard exponential distribution with cumulative distribution function $F_h(x)=1-e^{-x}$. Then, given ransom amount $r$, type $A_1$ hacker can obtain an expected payoff
\[
(\eta_1(r)-c_4)\mathbb{I}(\eta_1(r)>c_4),
\]
and type $A_2$ hacker can attain an expected payoff
\[
(\eta_2(r)-c_4)\mathbb{I}(\eta_2(r)>c_4).
\]

Figure \ref{fig04a} plots the ransom amount, victim type, and type $A_1$ hacker's equilibrium. The $x$-axis represents the ransom amount $r$, and the $y$-axis represents the victim's valuation on the files. The red vertical line is the equilibrium of type $A_1$ hacker, the black curve is the boundary curve separating victims' weakly dominant strategies.
When facing the equilibrium ransom claim from type $A_1$ hacker, the best strategy for the victim is as follows:
\begin{itemize}
\item[i)] if the  valuation of the file lies beneath the blue horizontal line, the victim should discard the files, i.e., $D$;
\item[ii)] if the  valuation of the file lies between the green line and the blue line, the victim should pay the ransom directly, i.e., $P$;
\item[iii)] if  the  valuation of the file lies above the green line, the victim should choose to crack first, i.e., $C$.
\end{itemize}

%

\noindent  Figure \ref{fig04b} plots the ransom amount, victim type, and type $A_2$ hacker's equilibrium, and we have similar observations. {From the two plots, we can see that the equilibrium ransom amount of type $A_2$ hacker is smaller than that of type $A_1$ hacker, which coincides with the finding of Theorem \ref{the05}(i). Moreover, when the victim is faced with type $A_1$ hacker, the threshold of valuation on the encrypted files such that the victim will choose to discard is higher, the threshold of valuation such that the victim will choose to crack is smaller, and the range of paying the ransom is smaller. This implies that there will be more victims to discard the files or crack when faced with type $A_1$ hacker, and to pay the ransom when faced with type $A_2$ hacker.}

\begin{figure}[!htbp]\centering
\subfigure[Equilibria: boundary curve (black curve), $S_h^e(A_1)$ (red dot)]
{\includegraphics[width=0.45\textwidth]{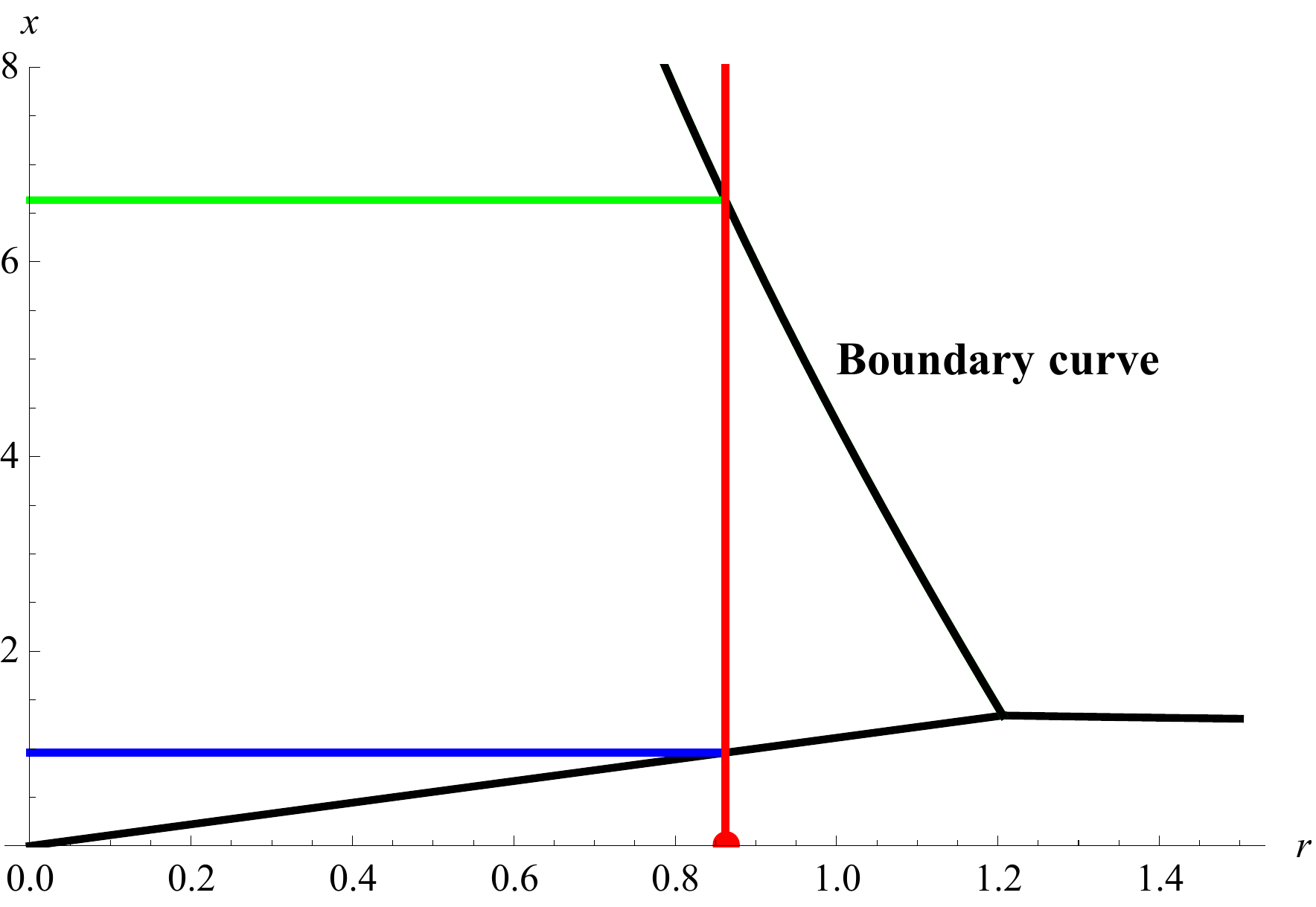}\label{fig04a}}
\subfigure[Equilibria: boundary curve (black curve), $S_h^e(A_2)$ (red dot)]
{\includegraphics[width=0.45\textwidth]{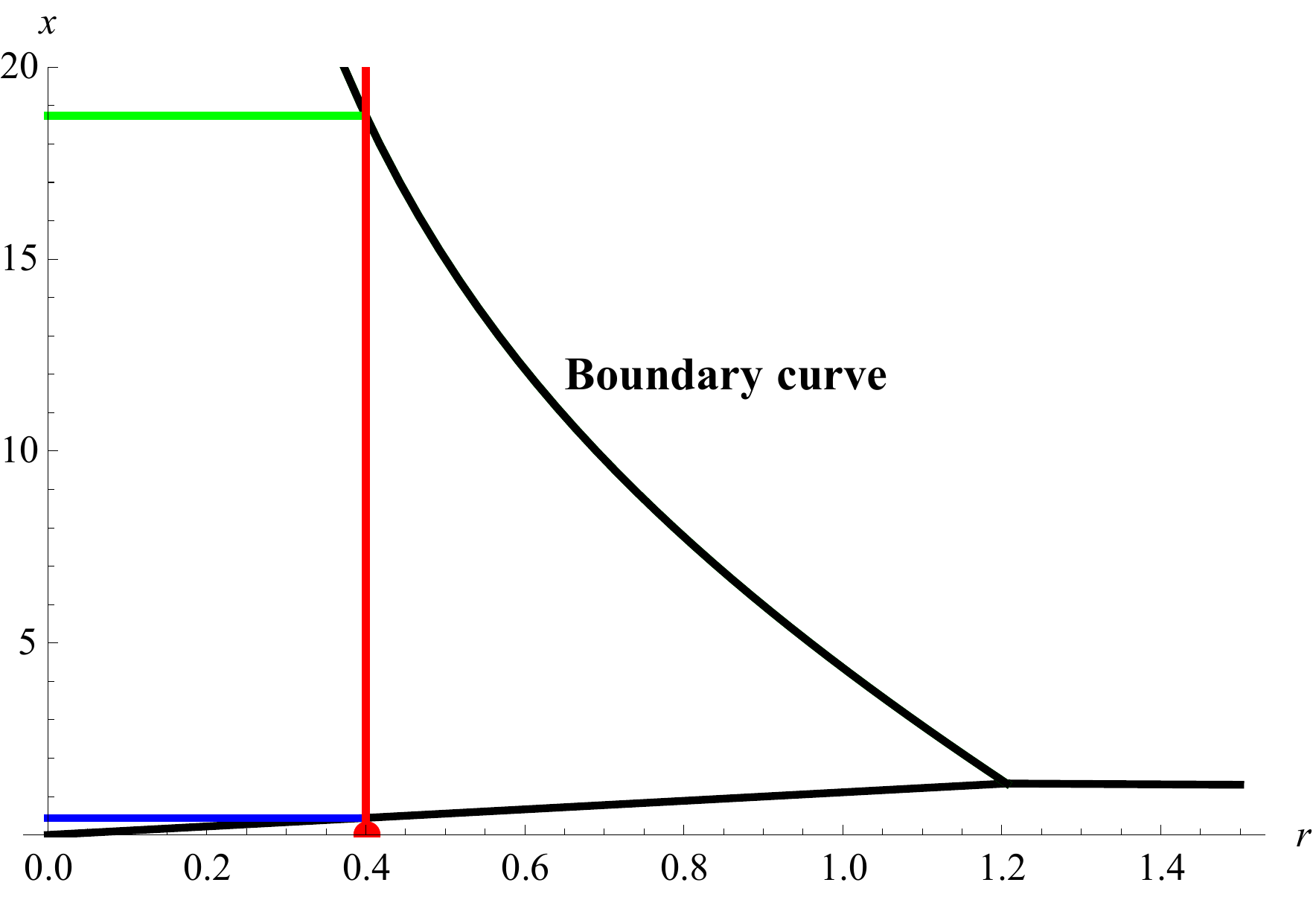}\label{fig04b}}
\caption{Equilibria of the Game $\Gamma_1$.}
\end{figure}

\end{Exa}

\subsection{Scenario II: victim has the backup}\label{case2}
In this section, we extend the model by allowing the option of recovering the encrypted files from the backup. In practice, institutions and individuals can back up important files regularly. With the backup, the recovery may not be a problem any longer. However, it may be very costly to regularly back up files.  Note that the backups are often stored in the network drives. Since almost all ransomwares encrypt the network drives connected to victims' computer, even victims back up files, there still exists the possibility that the encrypted files cannot be recovered.

We assume that if the victim decides not to pay the ransom directly, he/she now can first try to recover the files from the backup, and only when this action fails, the other actions  can be taken. That is, the victim's action set is given by {$\mathcal{A}_v=\{D,P,R,C\}$} 
where $R$ represents that the victim tries to recover the files from the backup first. Since the victim now has the recovery option,  we assume that the victim will try to crack the encrypted files only when the recovery is failed. We use $\Gamma_2$ as the strategic-form of the game \eqref{game} under this case. The costs and successful cracking probability are assumed to be the same as those   in Section \ref{case1}.

Before presenting the theorem, we first introduce the next lemma.

\begin{Lem}\rm\label{lem02}
Denote
\[
\psi_3(r)=\frac{[c_1+(c_2+r)(1-p_1)p_2(r)](1-p_3)+c_3-r}{(1-p)[1-(1-p_1)p_2(r)(1-p_3)]},
\]
and
\[
\psi_4(r)=\frac{[c_1+(c_2+r)(1-p_1)p_2(r)](1-p_3)+c_3}{1-(1-p_1)p_2(r)(1-p_3)(1-p)}.
\]
Let
\begin{eqnarray*}
& &\mathcal{M}_3=\big\{r\ge 0:r/p\le\psi_4(r)\le \psi_3(r)\big\},\quad\mathcal{M}_4=\big\{r\ge 0:\psi_3(r)\le\psi_4(r)\le r/p\big\}.
\end{eqnarray*}
Then $\mathcal{M}_1=[0,\omega_1]$ and $\mathcal{M}_2=[\omega_1,\infty)$, {where $\omega_1$ is the solution to the equation
\[
c_1p(1-p_3)+(c_2p+r)(1-p_1)p_2(r)(1-p_3)+c_3p-r=0.
\]}
\end{Lem}

{\bf Proof:} Note that
\begin{eqnarray*}
\psi_3(r)-\frac{r}{p}
&\stackrel{\text{sgn}}{=}&\psi_4(r)-\frac{r}{p}\\
&\stackrel{\text{sgn}}{=}&\psi_3(r)-\psi_4(r)\\
&\stackrel{\text{sgn}}{=}&c_1p(1-p_3)+(c_2p+r)(1-p_1)p_2(r)(1-p_3)+c_3p-r.
\end{eqnarray*}
The rest can be proved in a similar manner to that of Lemma \ref{lem00}. \qed

The following result shows the victim's weakly dominant strategy when facing a ransom demand from the hacker.
\begin{The}\rm\label{the03}
For the Bayesian game $\Gamma_2$, given the ransom $r$,
\begin{enumerate}[(i)]
\item if $r\in\mathcal{M}_1$, then it is weakly dominant for the victim to use the strategy
\[
S_v(x)=\left\{
         \begin{array}{ll}
           D, & x\le r/p, \\
           P, & r/p<x\le\psi_3(r),\\
           R, & x>\psi_3(r);
         \end{array}
       \right.
\]
\item if $r\in\mathcal{M}_2$, then it is weakly dominant for the victim to use the strategy
\[
S_v(x)=\left\{
         \begin{array}{ll}
           D, & x\le \psi_4(r), \\
           R, & x>\psi_4(r),
         \end{array}
       \right.
\]
and the victim should never pay the ransom first.
\end{enumerate}
\end{The}

{\bf Proof:} For a victim with private type $x\in[0,\infty)$, the utility functions can be represented as
\begin{eqnarray*}
& &U_v(x,A_1,D,r)=U_v(x,A_2,P,r)=-x,\\
& &U_v(x,A_1,P,r)=-r,\quad U_v(x,A_2,P,r)=-x-r,\\
& &U_v(x,A_1,R,r)=-c_3p_3-[c_1+(c_2+r)(1-p_1)p_2(r)+c_3](1-p_3)\\
& &=-c_3-[c_1+(c_2+r)(1-p_1)p_2(r)](1-p_3),\\
& &U_v(x,A_2,R,r)=-c_3p_3-[c_1+(x+c_2+r)(1-p_1)p_2(r)+c_3](1-p_3)\\
& &=-c_3-[c_1+(x+c_2+r)(1-p_1)p_2(r)](1-p_3).
\end{eqnarray*}

If the strategy $S_v(r)=D$ is used, then
\[
\mathsf{E}[U_v(x,T_h,D,r)]=-x.
\]
If the strategy $S_v(x)=P$ is used, then
\[
\mathsf{E}[U_v(x,T_h,P,r)]=U_v(x,A_1,P,r)p+U_v(x,A_2,P,r)(1-p)=-x(1-p)-r.
\]
If the strategy $S_v(x)=R$ is used, then
\begin{eqnarray*}
& &\mathsf{E}[U_v(x,T_h,R,r)]\\
&=&U_v(x,A_1,R,r)p+U_v(x,A_2,R,r)(1-p)\\
&=&[-c_3-(c_1+(c_2+r)(1-p_1)p_2(r))(1-p_3)]p\\
& &+[-c_3-(c_1+(x+c_2+r)(1-p_1)p_2(r))(1-p_3)](1-p)\\
&=&-x(1-p_1)p_2(r)(1-p_3)(1-p)-[c_1+(c_2+r)(1-p_1)p_2(r)](1-p_3)-c_3.
\end{eqnarray*}
Therefore, the difference between expected payoffs using pure strategies $D$ and $P$ is
\[
\mathsf{E}[U_v(x,T_h,D,r)]-\mathsf{E}[U_v(x,T_h,P,r)]\stackrel{\text{sgn}}{=}-x+r/p.
\]
The difference between expected payoffs using pure strategies $P$ and $R$ is
\[
\mathsf{E}[U_v(x,T_h,P,r)]-\mathsf{E}[U_v(x,T_h,R,r)]\stackrel{\text{sgn}}{=}-x+\psi_3(r).
\]
The difference between expected payoffs using pure strategies $D$ and $R$ is
\[
\mathsf{E}[U_v(x,T_h,D,r)]-\mathsf{E}[U_v(x,T_h,R,r)]\stackrel{\text{sgn}}{=}-x+\psi_4(r).
\]
 The rest of the proof can be completed by using the similar argument to that of Theorem \ref{the01}.  \qed

{\bf Remark:}  Theorem \ref{the03} shows that the victim can discard the files with a low valuation, and should try to recover first if the valuation is large. If the ransom amount is not that large and the files' valuation is medium, the victim can pay the ransom directly. {\em However, if the ransom amount is too large, the victim should never pay the ransom first}. Note that, for a victim with valuation $x$,  the weakly dominant strategy can be concisely expressed as
\[
D\mathbb{I}(x\le r/p\wedge \psi_4(r))+P\mathbb{I}(r/p<x\le \psi_3(r))+R\mathbb{I}(x>\psi_3(r)\vee \psi_4(r)).
\]

Note that the hacker's utility is affected by whether the victim pays the ransom. For type $A_1$ hacker, the utility functions are given by:
\[
U_h(x,A_1,D,r)=-c_4,\quad U_h(x,A_1,P,r)=r-c_4,\quad U_h(x,A_1,R,r)=rp_2(r)(1-p_1)(1-p_3)-c_4,
\]
and for the hacker with private type $A_2$, the utility functions are
\begin{eqnarray*}
U_h(x,A_2,D,r)&=&b_1-c_4,\\
U_h(x,A_2,P,r)&=&b_2+r-c_4,\\
U_h(x,A_2,R,r)&=&b_1(p_3+(1-p_3)p_1)+(b_2+rp_2(r))(1-p_1)(1-p_3)-c_4.
\end{eqnarray*}

In the following, we discuss the Bayesian Nash equilibrium of  game $\Gamma_2$. Assume $r_3^*$ is the smallest maximizer to
\begin{eqnarray*}
\eta_3(r)&=&rp_2(r)(1-p_1)(1-p_3)\big[\bar F_h(\psi_3(r))\mathbb{I}\big(r\in\mathcal{M}_3\big)+\bar F_h(\psi_4(r))\mathbb{I}\big(r\in\mathcal{M}_4\big)\big]\\
& &+r\big[\bar F_h(r/p)-\bar F_h(\psi_3(r))\big]\mathbb{I}\big(r\in\mathcal{M}_3\big),
\end{eqnarray*}
and $r_4^*$ is the smallest maximizer to
\begin{eqnarray*}
\eta_4(r)&=&[b_1(p_3+(1-p_3)p_1)+(b_2+rp_2(r))(1-p_1)(1-p_3)]\big[\bar F_h(\psi_3(r))\mathbb{I}\big(r\in\mathcal{M}_3\big)+\bar F_h(\psi_4(r))\mathbb{I}\big(r\in\mathcal{M}_4\big)\big]\\
& &+(b_2+r)\big[\bar F_h(r/p)-\bar F_h(\psi_3(r))\big]\mathbb{I}\big(r\in\mathcal{M}_3\big)\\
& &+b_1\big[F_h(r/p)\mathbb{I}\big(r\in\mathcal{M}_3\big)+F_h(\psi_4(r))\mathbb{I}\big(r\in\mathcal{M}_4\big)\big].
\end{eqnarray*}
By employing the similar proof argument to that of Lemma \ref{lem01}, we can verify the existences of $r_3^*$ and $r_4^*$ when $p_2(r)$ satisfies condition \eqref{con1}.

The next theorem shows the existence of pure Bayesian Nash equilibrium of game $\Gamma_2$.

\begin{The}\rm\label{the04}
Assume $p_2(\cdot)$ satisfies condition \eqref{con1}. For the Bayesian game $\Gamma_2$, there exists a pure Bayesian equilibrium such that the hacker and the victim use the pure strategy profile $\Big(S_h^e(A_i),S_v^e\big(\cdot\mid S_h^e(A_i)\big)\Big)$, where
\[
S_h^e(A_1)=r_3^*\mathbb{I}\big(\eta_3(r_3^*)>c_4\big),\quad\mbox{and}\quad S_h^e(A_2)=r_4^*\mathbb{I}\big(\eta_4(r_4^*)>c_4\big),
\]
and for any $x\in[0,\infty)$,
\begin{eqnarray}\label{eq10}
& &S_v^e\big(x\mid S_h^e(\cdot)\big)\\
&=&D\Big[\mathbb{I}\Big(x\le S_h^e(\cdot)/p\wedge\psi_4\big(S_h^e(\cdot)\big)\Big)\Big]+P\mathbb{I}\Big(S_h^e(\cdot)/p\le x\le \psi_3\big(S_h^e(\cdot)\big)\Big)\notag\\
& &+R\Big[\mathbb{I}\Big(x>\psi_3\big(S_h^e(\cdot)\big)\vee\psi_4\big(S_h^e(\cdot)\big)\Big)\Big].\notag
\end{eqnarray}
\end{The}

{\bf Proof:} Note that, for type $A_1$ hacker,
\begin{eqnarray*}
& &\mathsf{E}\big[U_h\big(V,A_1,S_v(V),r\big)\big]\\
&=&\sum_{i=3}^4\mathsf{E}\big[U_h\big(V,A_1,S_v(V),r\big)\big]\mathbb{I}\big(r\in\mathcal{M}_i\big)\\
&=&\eta_3(r)-c_4.
\end{eqnarray*}
Similarly,  for type $A_2$ hacker, we have
\begin{eqnarray*}
& &\mathsf{E}\big[U_h\big(V,A_2,S_v(V),r\big)\big]\\
&=&\sum_{i=3}^4\mathsf{E}\big[U_h\big(V,A_2,S_v(V),r\big)\big]\mathbb{I}\big(r\in\mathcal{M}_i\big)\\
&=&\eta_4(r)-c_4.
\end{eqnarray*}
The rest of the proof is similar to that of Theorem \ref{the02}, which is omitted for briefness. \qed

{\bf Remark:} Similar to Theorem \ref{the02}, Theorem \ref{the04} shows that in the equilibrium the attacker should have enough incentive to launch the ransomware attack. {\em If the attacker asks for a `correct' ransom amount, the victim may choose to directly pay the ransom even though the victim has the backup}. Further, type $A_2$ hacker always attains a larger expected payoff than type $A_1$ hacker does, because for all $r\ge 0$,
\begin{eqnarray*}
\eta_4(r)-\eta_3(r)&=&[b_1(p_3+(1-p_3)p_1)+b_2(1-p_1)(1-p_3)]\big[\bar F_h(\psi_3(r))\mathbb{I}\big(r\in\mathcal{M}_3\big)+\bar F_h(\psi_4(r))\mathbb{I}\big(r\in\mathcal{M}_4\big)\big]\\
& &+b_2\big[\bar F_h(r/p)-\bar F_h(\psi_3(r))\big]\mathbb{I}\big(r\in\mathcal{M}_3\big)\\
& &+b_1\big[F_h(r/p)\mathbb{I}\big(r\in\mathcal{M}_3\big)+F_h(\psi_4(r))\mathbb{I}\big(r\in\mathcal{M}_4\big)\big]\\
&\ge&0.
\end{eqnarray*}

 In a similar manner to Theorem \ref{the05}, we can prove a similar relation between the equilibrium ransoms of the type $A_1$ hacker and type $A_2$ hacker in game $\Gamma_2$. The proof is omitted for brevity.

\begin{The}\rm\label{the06} For the Bayesian game $\Gamma_2$ in equilibrium,  assume $rp_2(\cdot)$ is decreasing, and  the equilibrium ransom amounts of hacker $A_1$ and $A_2$ are $r_3^*$ and $r_4^*$, respectively. Then,
\begin{enumerate}[(i)]
\item If $r_4^*\in\mathcal{M}_3$, then $r_3^*\ge r_4^*$;
\item If $r_4^*\in\mathcal{M}_4$, then $r_3^*\le r_4^*$.
\end{enumerate}
\end{The}

Similarly to Theorem \ref{cor01}, the following result shows that there exists a randomized Bayesian Nash equilibrium for the game $\Gamma_2$ when the maximizers to $\eta_3$ and $\eta_4$ are not unique, respectively.

\begin{The}\rm\label{cor02}
Assume $p_2(\cdot)$ fulfills condition \eqref{con1}. For the Bayesian game $\Gamma_2$, if $\mathcal{R}_3=\mathop{\arg\max}_r\eta_3(r)$ and $\mathcal{R}_4=\mathop{\arg\max}_r\eta_4(r)$ are both non-singleton, then the randomized Bayesian Nash equilibrium of the game is given by $\Big(S_h^e(A_i),S_v^e\big(\cdot\mid S_h^e(A_i)\big)\Big)$, with
\[
S_h^e(A_1)=\sigma_3\mathbb{I}\big(\eta_3(r_3^*)>c_4\big),\quad\mbox{and}\quad S_h^e(A_2)=\sigma_4\mathbb{I}\big(\eta_4(r_4^*)>c_4\big),
\]
and $S_v^e\big(x\mid S_h^e(\cdot)\big)$ is given by \eqref{eq10}, where
\[
\bigg\{\sigma_{i}:\mathcal{R}_i\mapsto [0,1]:\sum_{r\in\mathcal{R}_{i}}\sigma_{i}(r)=1\bigg\},
\]
for $i=3,4$.
\end{The}

In the following example, we illustrate the best response of the victim given any ransom demand and the pure equilibria of both types of hackers in  game $\Gamma_2$.

\begin{Exa}\rm\label{exa04}
Assume that in  game $\Gamma_2$, $c_3=0.2$, $p_1=0.1$, $p_3=0.3$, and the other parameters are the same as those in Example \ref{exa03}. Then, given ransom amount $r$, type $A_1$ hacker can obtain an expected payoff
\[
(\eta_3(r)-c_4)\mathbb{I}(\eta_3(r)>c_4),
\]
and type $A_2$ hacker can attain an expected payoff
\[
(\eta_4(r)-c_4)\mathbb{I}(\eta_4(r)>c_4).
\]

Figure \ref{fig04c} shows the victim's best strategy  when facing type $A_1$ hacker. The $x$-axis represents the ransom amount $r$, and the $y$-axis represents the victim's valuation on the files. The red  vertical line is the equilibrium of type $A_1$ hacker, the black curve is the boundary curve separating victims' weakly dominant strategies.
When facing the equilibrium ransom claim from type $A_1$ hacker, the best strategy for the victim is as follows:
\begin{itemize}
\item[i)] if the  valuation of the file lies beneath the blue horizontal line, the victim should discard the files, i.e., $D$;
\item[ii)] if the  valuation of the file lies between the green line and the blue line, the victim should pay the ransom directly, i.e., $P$;
\item[iii)] if  the  valuation of the file lies above the green line, the victim should choose to recover first, i.e., $R$.
\end{itemize}

{ Figure \ref{fig04d} shows the victim's best strategy when facing type $A_2$ hacker, and similar explanation to Figure \ref{fig04c} holds for the type $A_2$ hacker in Figure \ref{fig04d}.} From the two plots, we can observe similar relations as in the two plots of Example \ref{exa03}. Specifically, the equilibrium ransom amount of type $A_2$ hacker is smaller than that of type $A_1$ hacker, which coincides with the finding of Theorem \ref{the06}(i). Moreover, when the victim is faced with type $A_1$ hacker, the threshold of valuation on the encrypted files such that the victim will choose to discard is higher, the threshold of valuation such that the victim will choose to recover is smaller, and the range of paying the ransom is smaller. This implies that there will be more victims to discard the files or try to recover when facing type $A_1$ hacker, and to pay the ransom when facing type $A_2$ hacker.

\begin{figure}[!htbp]\centering
\subfigure[Equilibria: boundary curve (black curve), $S_h^e(A_1)$ (red dot)]
{\includegraphics[width=0.45\textwidth]{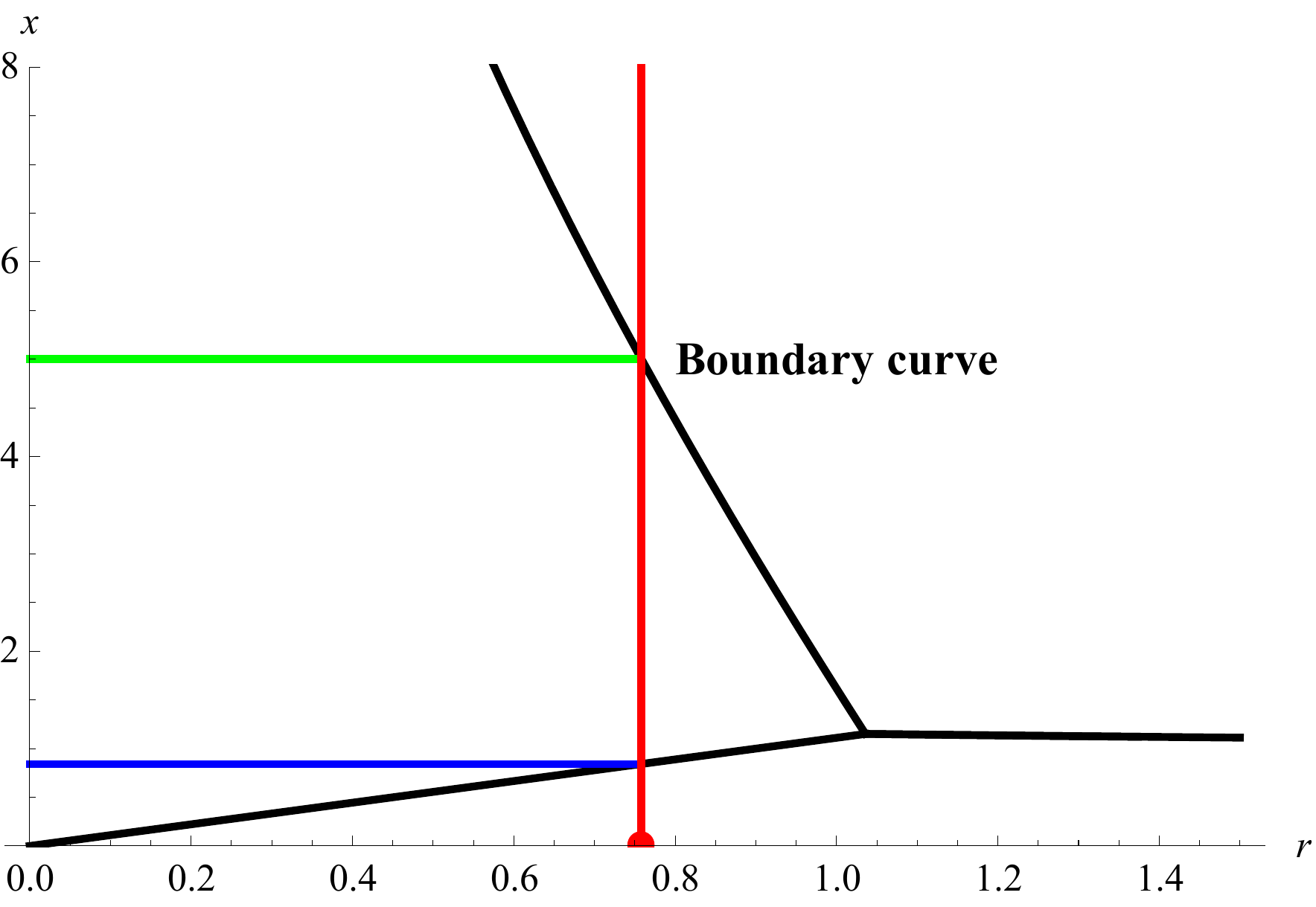}\label{fig04c}}
\subfigure[Equilibria: boundary curve (black curve), $S_h^e(A_2)$ (red dot)]
{\includegraphics[width=0.45\textwidth]{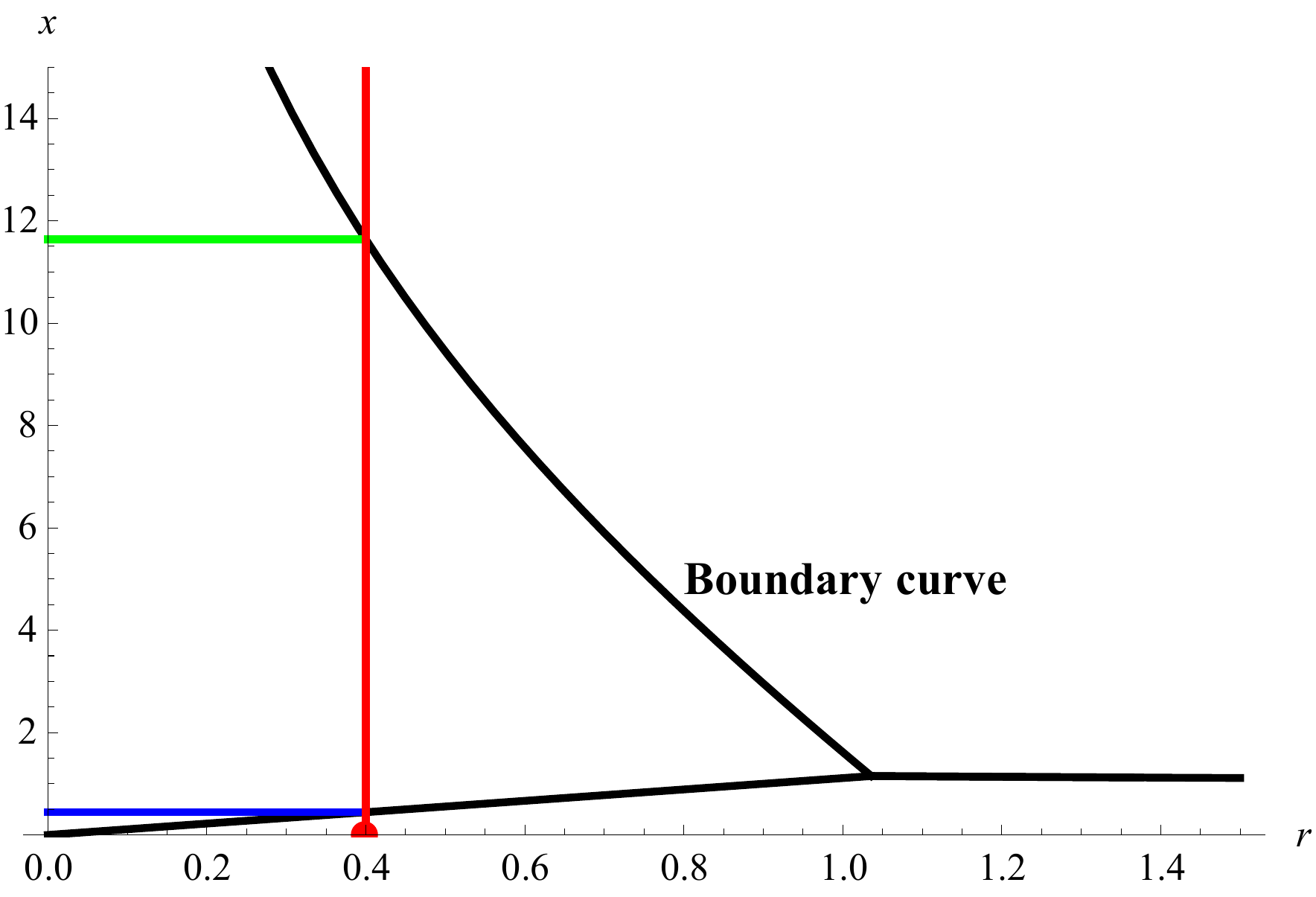}\label{fig04d}}
\caption{Equilibria of the Game $\Gamma_2$.}
\end{figure}

\end{Exa}

\section{Comparative study and  examples}\label{sec:exm}
In this section, we investigate the effects of different parameters on the equilibria  of both  hacker and  victim. For the illustration purpose, we  consider the pure Bayesian Nash equilibria of the two games studied in the previous sections. With the same parameter configuration, we will also compare the hacker's equilibrium strategies under the frameworks without backup and with backup to illustrate the effect of the recovery protection procedure.

We first study how the costs $(c_1,c_2,c_3,c_4)$  affect the expected payoffs of both types of hackers in both games $\Gamma_1$ and $\Gamma_2$.
\begin{Pro}\rm\label{pro01}
\begin{enumerate}[(i)]
\item For game $\Gamma_1$, the expected payoffs of both types of hackers
\begin{itemize}
\item[a)] increase in $c_1$ and $c_2$, if $r\in\mathcal{M}_1$;
\item[b)] decrease in $c_1$ and $c_2$, if $r\in\mathcal{M}_2$; and
\item[c)] decrease in $c_4$ for all $r$.
\end{itemize}

\item For game $\Gamma_2$, the expected payoffs of both types of hackers
\begin{itemize}
\item[a)] increase in $c_1$, $c_2$, $c_3$ if $r\in\mathcal{M}_1$;
\item[b)] decrease in $c_1$, $c_2$, $c_3$ if $r\in\mathcal{M}_2$; and
\item[c)] decrease in $c_4$ for all $r$.
\end{itemize}
\end{enumerate}
\end{Pro}

{\bf Proof:} We only prove the case (i), and the case (ii) case can be proved similarly.  From Eq. \eqref{eq02}, the
expected payoff  of type $A_1$ attacker is
\begin{eqnarray*}
\mathsf{E}\big[U_h\big(V,A_1,S_v(V),r\big)\big]&=&rp_2(r)(1-p_1)\big[\bar F_h(\psi_1(r))\mathbb{I}\big(r\in\mathcal{M}_1\big)+\bar F_h(\psi_2(r))\mathbb{I}\big(r\in\mathcal{M}_2\big)\big] \\
&& +r\big[\bar F_h(r/p)-\bar F_h(\psi_1(r))\big]\mathbb{I}\big(r\in\mathcal{M}_1\big) -c_4.
\end{eqnarray*}
{ According to Eqs. \eqref{eq:psi1} and \eqref{eq:psi2}, $\psi_1(r)$ and $\psi_2(r)$ increase in $c_1$ and $c_2$, which implies that $\bar F_h(\psi_1(r))$ and $\bar F_h(\psi_2(r))$ decrease in $c_1$ and $c_2$.} Therefore, the
expected payoff of type $A_1$ attacker is increases in $c_1$ and $c_2$ if $r\in\mathcal{M}_1$, and decreasing in $c_1$ and $c_2$ if $r\in\mathcal{M}_2$. The expected payoff of either type of hacker clearly deceases in $c_4$.

This completes the proof.  \qed

{\bf Remark:} Note that $c_1$ (i.e., the cost of cracking files)
 and $c_2$ (i.e., the extra cost upon the failure of cracking files) measure the powerfulness of the victim in the combat with the hacker. A larger value of $c_1$ or $c_2$ usually means the victim is weaker in the sense that the victim has less ability to deal with the situation of ransomware attack. At the first glance, the hacker may be able to exploit this weakness of the victim and attain a larger expected payoff in any situation. However, Theorems \ref{the01} and \ref{the03} reveal that if the ransom amount is very large,
the victim may discard the files directly. This action is actually adverse to the hacker's payoff. Therefore, the monotonicity of expected payoff of the hacker on $c_1$ or $c_2$ depends on the demanded ransom amount. The parameter $c_4$ is the hacker's cost on planning and launching the attack, and hence  naturally has a negative effect on the hacker's expected payoffs.

In the following, we discuss the impact of the event probabilities on the hackers' expected payoffs including  $p_1$ (i.e., successfully  cracking the encrypted files), $p_2(r)$ (paying the ransom with an extra punishment fee $c_2$), and $p_3$ (successfully recovering from the backup).   Due to the different reactions of type $A_1$ and $A_2$ hackers after receiving the ransom, the probability of encountering type $A_1$ hacker has some potential effect on the victims' strategies, and hence the hackers' expected payoffs. Therefore, we also study the impact of $p$ (i.e., encountering type $A_1$ hacker).

{
\begin{Pro}\rm\label{pro04} Assume that both games are in equilibria.
\begin{itemize}
\item[i)] For game $\Gamma_1$, denote the equilibrium ransom amounts of hacker $A_1$ and $A_2$ as $r_1^*$ and $r_2^*$, respectively. Then
    \begin{itemize}
    \item[a)] the expected payoff of hacker $A_1$ decreases (increases) in $p_1$, and increases (decreases) in $p_2$ if $r_1^*\in\mathcal{M}_1$ ($r_1^*\in\mathcal{M}_2$);
    \item[b)] the expected payoff of hacker $A_2$ decreases (increases) in $p_1$, and increases (decreases) in $p_2$ if $r_2^*\in\mathcal{M}_1$ ($r_2^*\in\mathcal{M}_2$);
    \item[c)] the expected payoffs of both types of hackers increase in $p$.
    \end{itemize}
\item[ii)] For game $\Gamma_2$, denote the equilibrium ransom amounts of hacker $A_1$ and $A_2$ as $r_3^*$ and $r_4^*$, respectively. Then
    \begin{itemize}
    \item[a)] the expected payoff of hacker $A_1$ decreases (increases) in $p_1$ and $p_3$, and increases (decreases) in $p_2$ if $r_3^*\in\mathcal{M}_3$ ($r_4^*\in\mathcal{M}_4$);
    \item[b)] the expected payoff of hacker $A_2$ decreases (increases) in $p_1$ and $p_3$, and increases (decreases) in $p_2$ if $r_3^*\in\mathcal{M}_3$ ($r_4^*\in\mathcal{M}_4$);
    \item[c)] the expected payoffs of both types of hackers increase in $p$.
    \end{itemize}
\end{itemize}
\end{Pro}
}

{\bf Proof:} We only prove the case of i) as the case of ii) can be proved in a similar manner.

Denote \[
\psi_0(x;a,b)=\frac{a+bx}{1-x}
\]
with $b\ge 0$. By using the function $\psi_0(x;a,b)$, the functions $\psi_1(r)$ in Eq. \eqref{eq:psi1} and $\psi_2(r)$ in Eq. \eqref{eq:psi2}  can be equivalently represented as
\[
\psi_1(r)=\frac{c_1-r+(c_2+r)p_2(r)(1-p_1)}{(1-p)[1-p_2(r)(1-p_1)]}
=\psi_0\bigg(p_2(r)(1-p_1);\frac{c_1-r}{1-p},\frac{c_2+r}{1-p}\bigg),
\]
and
\[
\psi_2(r)=\frac{c_1+(c_2+r)p_2(r)(1-p_1)}{1-p_2(r)(1-p_1)(1-p)}
=\psi_0\bigg(p_2(r)(1-p_1)(1-p);c_1,\frac{c_2+r}{1-p}\bigg).
\]
Since $\psi_0(x;a,b)$ is increasing in $x\in(0,1)$ for $a+bx\ge 0$,
it holds that for $r$ such that $\psi_1(r)\ge 0$, namely,
\[
r\le \frac{c_1+c_2p_2(r)(1-p_1)}{1-p_2(r)(1-p_1)},
\]
{$p_1'\le p_1$ implies
\[
\psi_0\bigg(p_2(r)(1-p_1');\frac{c_1-r}{1-p},\frac{c_2+r}{1-p}\bigg)\ge \psi_0\bigg(p_2(r)(1-p_1);\frac{c_1-r}{1-p},\frac{c_2+r}{1-p}\bigg),
\]
and $p_2'(r)\ge p_2(r)$ implies
\[
\psi_0\bigg(p_2'(r)(1-p_1);\frac{c_1-r}{1-p},\frac{c_2+r}{1-p}\bigg)\ge \psi_0\bigg(p_2(r)(1-p_1);\frac{c_1-r}{1-p},\frac{c_2+r}{1-p}\bigg).
\]
Also, one can verify that $p'\le p$ implies
\[
\psi_0\bigg(p_2(r)(1-p_1);\frac{c_1-r}{1-p'},\frac{c_2+r}{1-p'}\bigg)\le \psi_0\bigg(p_2(r)(1-p_1);\frac{c_1-r}{1-p},\frac{c_2+r}{1-p}\bigg).
\]
Similarly, concerning $\psi_2(r)$, it follows that
\[
p_1'\le p_1\Rightarrow \psi_0\bigg(p_2(r)(1-p_1')(1-p);c_1,\frac{c_2+r}{1-p}\bigg)\ge \psi_0\bigg(p_2(r)(1-p_1)(1-p);c_1,\frac{c_2+r}{1-p}\bigg),
\]
\[
p_2'(r)\ge p_2(r)\Rightarrow \psi_0\bigg(p_2'(r)(1-p_1)(1-p);c_1,\frac{c_2+r}{1-p}\bigg)\ge\psi_0\bigg(p_2(r)(1-p_1)(1-p);c_1,\frac{c_2+r}{1-p}\bigg),
\]
and
\[
p'\le p\Rightarrow \psi_0\bigg(p_2(r)(1-p_1)(1-p');c_1,\frac{c_2+r}{1-p'}\bigg)\ge\psi_0\bigg(p_2(r)(1-p_1)(1-p);c_1,\frac{c_2+r}{1-p}\bigg)
\]
Therefore, for $r_1^*\in\mathcal{M}_1$, the expected payoff of type $A_1$ hacker with a ransom amount $r$ is given by
\[
\eta_1(r;p_2(r),p_1,p)-c_4=r[p_2(r)(1-p_1)-1]\bar F_h(\psi_1(r))+r\bar F_h(r/p)-c_4,
\]
which satisfies that
\[
p_1'\le p_1\Longrightarrow \eta_1(r_1^*;p_2(r),p_1',p)\ge \eta_1(r_1^*;p_2(r),p_1,p),
\]
\[
p_2'(r)\ge p_2(r)\Longrightarrow \eta_1(r_1^*;p_2'(r),p_1,p)\ge \eta_1(r_1^*;p_2(r),p_1,p),
\]
and
\[
p'\le p\Longrightarrow \eta_1(r_1^*;p_2(r),p_1,p')\le \eta_1(r_1^*;p_2(r),p_1,p).
\]
As for $r_1^*\in\mathcal{M}_2$, the expected payoff of type $A_1$ hacker with a ransom amount $r$ given by
\[
\eta_1(r;p_2(r),p_1,p)-c_4=rp_2(r)(1-p_1)\bar F_h(\psi_2(r))-c_4,
\]
satisfying that
\[
p_1'\le p_1\Longrightarrow \eta_1(r_1^*;p_2(r),p_1',p)\le \eta_1(r_1^*;p_2(r),p_1,p),
\]
\[
p_2'(r)\ge p_2(r)\Longrightarrow \eta_1(r_1^*;p_2'(r),p_1,p)\le \eta_1(r_1^*;p_2(r),p_1,p),
\]
and
\[
p'\le p\Longrightarrow \eta_1(r_1^*;p_2(r),p_1,p')\le \eta_1(r_1^*;p_2(r),p_1,p).
\]

The case of type $A_2$ hacker can be proved similarly, and hence the proof is omitted.}  \qed

{ {\bf Remark:} The intuition behind the findings of Proposition \ref{pro04} lies in the fact that the hacker and the victim compete each other for the files. Hence, enhancing the probability of events benefiting the victim will naturally harm the hacker's benefit. Consider the game $\Gamma_1$. When the ransom amount asked by the hacker is relatively small, victims having a larger probability of $p_1$ or a smaller probability of $p_2(r)$ are more likely to crack the files rather than to pay the ransom, and hence the expected payoffs of the hackers are smaller. When the ransom amount is relatively large, victims having a larger probability of $p_1$ or a smaller probability of $p_2(r)$ are more likely to crack the files rather than to discard the files, and hence the expected payoffs of the hackers are larger. Concerning the probability $p$ of encountering a type $A_1$ hacker, since a type $A_2$ hacker will not decrypt the files even after receiving the ransom, victims may somewhat prefer to encounter a type $A_1$ hacker. Further, according to Theorems \ref{the01} and \ref{the03}, in the game $\Gamma_1$, if victims believe that there are more type $A_1$ hackers, they will be more likely to pay a relatively small ransom or to crack the file when facing a relatively large ransom amount, and the possibility of choosing to discard the files will be lower. This results in higher expected payoffs for either types of hackers. Similar intuition holds for the hackers and victims in the game $\Gamma_2$.}

\medskip
In the following example, we illustrate the monotone property with  respect to $p_1$ in Proposition \ref{pro04}.

\begin{Exa}\rm\label{exa01}
Consider the game $\Gamma_1$ with parameter configuration in Example \ref{exa03}. Figures \ref{fig01a} and \ref{fig01b} plot the curves of type $A_1$ and type $A_2$ hackers' expected payoffs with $p_1=0.1$ and $p_1=0.3$ versus the transformed ransom amount $r$, respectively\footnote{To display the whole of expected payoff curves on $[0,+\infty)$, we perform the transformation $(r+1)^{-1}$ for $r\in[0,+\infty)$ projecting $[0,+\infty)$ onto $[0,1]$.}. It is seen that in each plot, the curve of expected payoffs with $p_1=0.3$ is under that of the curve with $p_1=0.1$, confirming the finding of Proposition \ref{pro04}.

\begin{figure}[!htbp]\centering
\subfigure[ $A_1$: $p_1=0.1$ (black, solid), $p_1=0.3$ (red, dashed)]
{\includegraphics[height=0.2\textheight,width=0.475\textwidth]{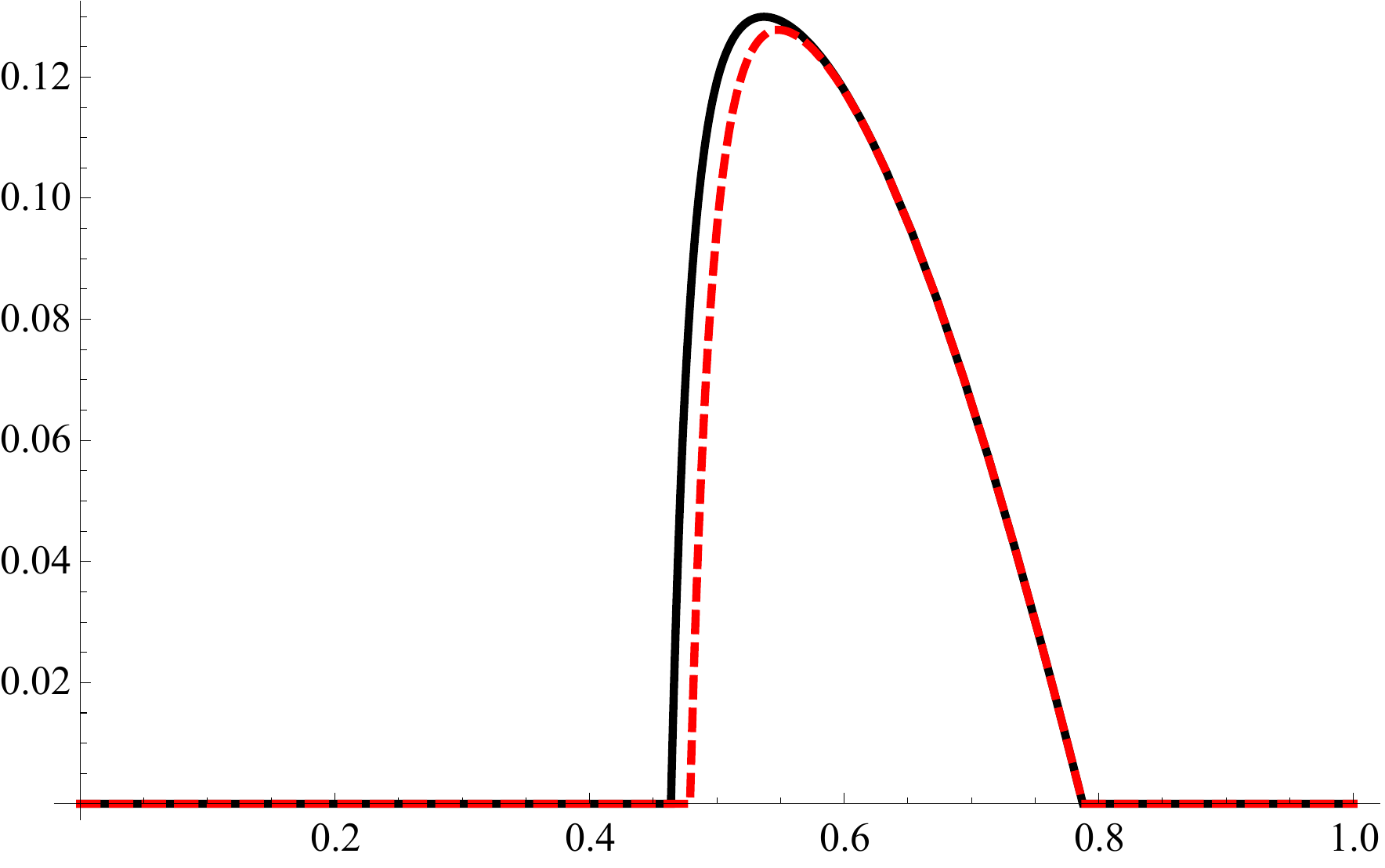}\label{fig01a}}
\subfigure[$A_2$: $p_1=0.1$ (black, solid), $p_1=0.3$ (red, dashed)]
{\includegraphics[height=0.23\textheight,width=0.475\textwidth]{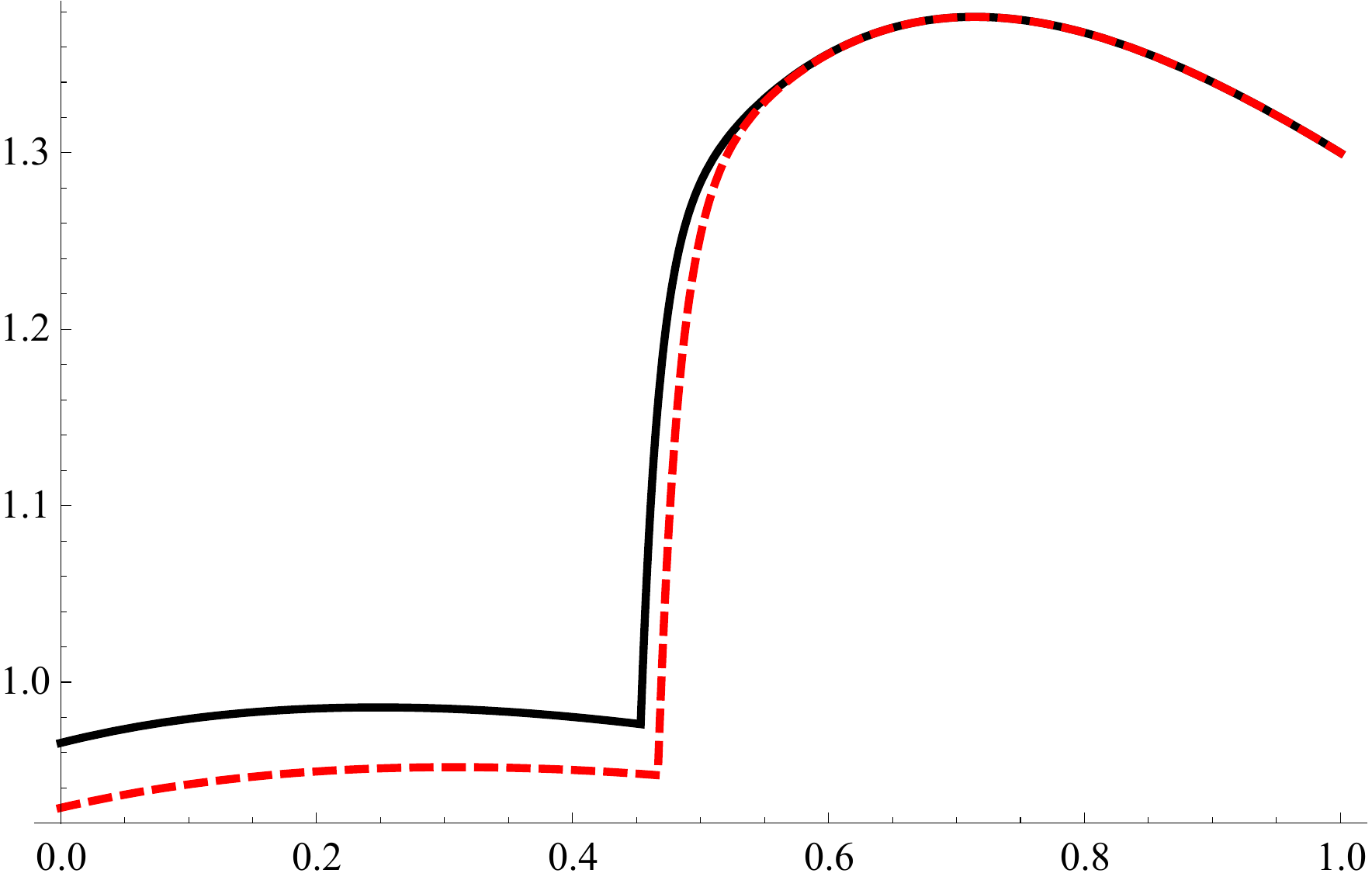}\label{fig01b}}
\caption{Expected payoffs of hackers versus the transformed ransom amount  $(r+1)^{-1}$.}
\end{figure}
\end{Exa}

In the following, we compare the hacker's expected payoffs in the games without the backup and with the backup given the same parameter configuration.

\begin{Pro}\rm\label{pro06}
Under the same parameter configuration, if $c_3\le p_3c_1$,
the expected payoffs of  both types of hackers in game $\Gamma_2$ are smaller than those in game $\Gamma_1$ for any ransom amount $r\in\mathcal{M}_3$, respectively.
\end{Pro}

{\bf Proof:} We only verify the case of type $A_1$ hacker and the other case can be proved similarly. Note that  if $c_3\le p_3c_1$, 
then for $r$ such that $\psi_3(r)\ge 0$, it holds that
\begin{eqnarray*}
\psi_3(r)
&=&\frac{c_1-c_1p_3+c_3+(c_2+r)(1-p_1)p_2(r)(1-p_3)-r}{(1-p)[1-(1-p_1)p_2(r)(1-p_3)]}\\
&\le&\frac{c_1-r+(c_2+r)(1-p_1)p_2(r)(1-p_3)}{[1-(1-p_1)p_2(r)(1-p_3)](1-p)}\\
&=&\psi_0\bigg(p_2(r)(1-p_1)(1-p_3);\frac{c_1-r}{1-p},\frac{c_2+r}{1-p}\bigg)\\
&\le&\psi_0\bigg(p_2(r)(1-p_1);\frac{c_1-r}{1-p},\frac{c_2+r}{1-p}\bigg)\\
&=&\psi_1(r).
\end{eqnarray*}
Therefore, it follows that
\begin{equation}\label{eq:g1}
\bar F_h(\psi_3(r))\ge\bar F_h(\psi_1(r))
\end{equation}
 for all $r\ge 0$.
{ Note that, for $r\in\mathcal{M}_3$, it holds that
\begin{eqnarray*}
0&\le&\psi_3(r)-\psi_4(r)\\
&\stackrel{\text{sgn}}{=}&c_1p(1-p_3)+c_2pp_2(r)(1-p_1)(1-p_3)+rp_2(r)(1-p_1)(1-p_3)-r\\
&=&(1-p_3)[c_1p+c_2pp_2(r)(1-p_1)+rp_2(r)(1-p_1)-r]-rp_3.
\end{eqnarray*}
Thus,  we have, for $r\in\mathcal{M}_3$,
\begin{equation}\label{eq:g2}
\psi_1(r)-\psi_2(r)\stackrel{\text{sgn}}{=}c_1p+c_2pp_2(r)(1-p_1)+rp_2(r)(1-p_1)-r\ge 0,
\end{equation}}
which implies that $r\in\mathcal{M}_1$ as well.
Note that given ransom $r$, type $A_1$ hacker expects to earn
\[
(\eta_1(r)-c_4)\mathbb{I}(\eta_1(r)>c_4)
\]
in the game $\Gamma_1$, and earn
\[
(\eta_3(r)-c_4)\mathbb{I}(\eta_3(r)>c_4)
\]
in the game $\Gamma_2$.
Further, for the type $A_1$ hacker, we have
\[
\eta_1(r)=-r[1-p_2(r)(1-p_1)]\bar F_h(\psi_1(r))+r\bar F_h(r/p)
\]
in the game $\Gamma_1$ for $r\in\mathcal{M}_3$ based on Eq. \eqref{eq:g2}, and
\[
\eta_3(r)=-r[1-p_2(r)(1-p_1)(1-p_3)]\bar F_h(\psi_3(r))+r\bar F_h(r/p)
\]
in the game $\Gamma_2$. Therefore, the desired result follows from Eq. \eqref{eq:g1}.   \qed

{\bf Remark:} The condition $c_3\le p_3c_1$ means that the ratio of the cost of recovering from the backup to that of cracking is not larger than the successful probability of recovering. That is, when the cost of recovery is relatively small compared to the cost of cracking,  both type $A_1$ and  $A_2$ hackers tend to earn less in game $\Gamma_2$ for ransom amount $r$ that is not too large.

\begin{Exa}[Example \ref{exa01} continued]\rm\label{exa02}
Consider two games $\Gamma_1$ and $\Gamma_2$ with the same setting in Example \ref{exa01}. Assume $c_3=0.2$, $p_1=0.1$ and $p_3=0.3$. Then, in  game $\Gamma_1$, given ransom amount $r$, type $A_1$ hacker can obtain an expected payoff
\[
(\eta_1(r)-c_4)\mathbb{I}(\eta_1(r)>c_4),
\]
and type $A_2$ hacker can attain an expected payoff
\[
(\eta_2(r)-c_4)\mathbb{I}(\eta_2(r)>c_4).
\]
In   game $\Gamma_2$, given ransom amount $r$, type $A_1$ hacker can obtain an expected payoff
\[
(\eta_3(r)-c_4)\mathbb{I}(\eta_3(r)>c_4),
\]
and type $A_2$ hacker can attain an expected payoff
\[
(\eta_4(r)-c_4)\mathbb{I}(\eta_4(r)>c_4).
\]

Figures \ref{fig02a} and \ref{fig02b} plot the curves of type $A_1$ and type $A_2$ hackers' expected payoffs  in the games $\Gamma_1$ and $\Gamma_2$ versus the transformed ransom amount  $(r+1)^{-1}$, respectively. It is seen that the curves of expected payoffs in $\Gamma_2$ are under those in $\Gamma_1$ in both plots for all ransom amount, confirming the finding of Proposition \ref{pro06}. Moreover, it can also be seen that for large ransom amounts (corresponding to small values of the x-axis in the plots), the hackers' expected payoffs in $\Gamma_2$ are also smaller than those in $\Gamma_1$.

\begin{figure}[!htbp]\centering
\subfigure[$A_1$: $\Gamma_1$ (black, solid), $\Gamma_2$ (red, dashed)]
{\includegraphics[height=0.2\textheight,width=0.475\textwidth]{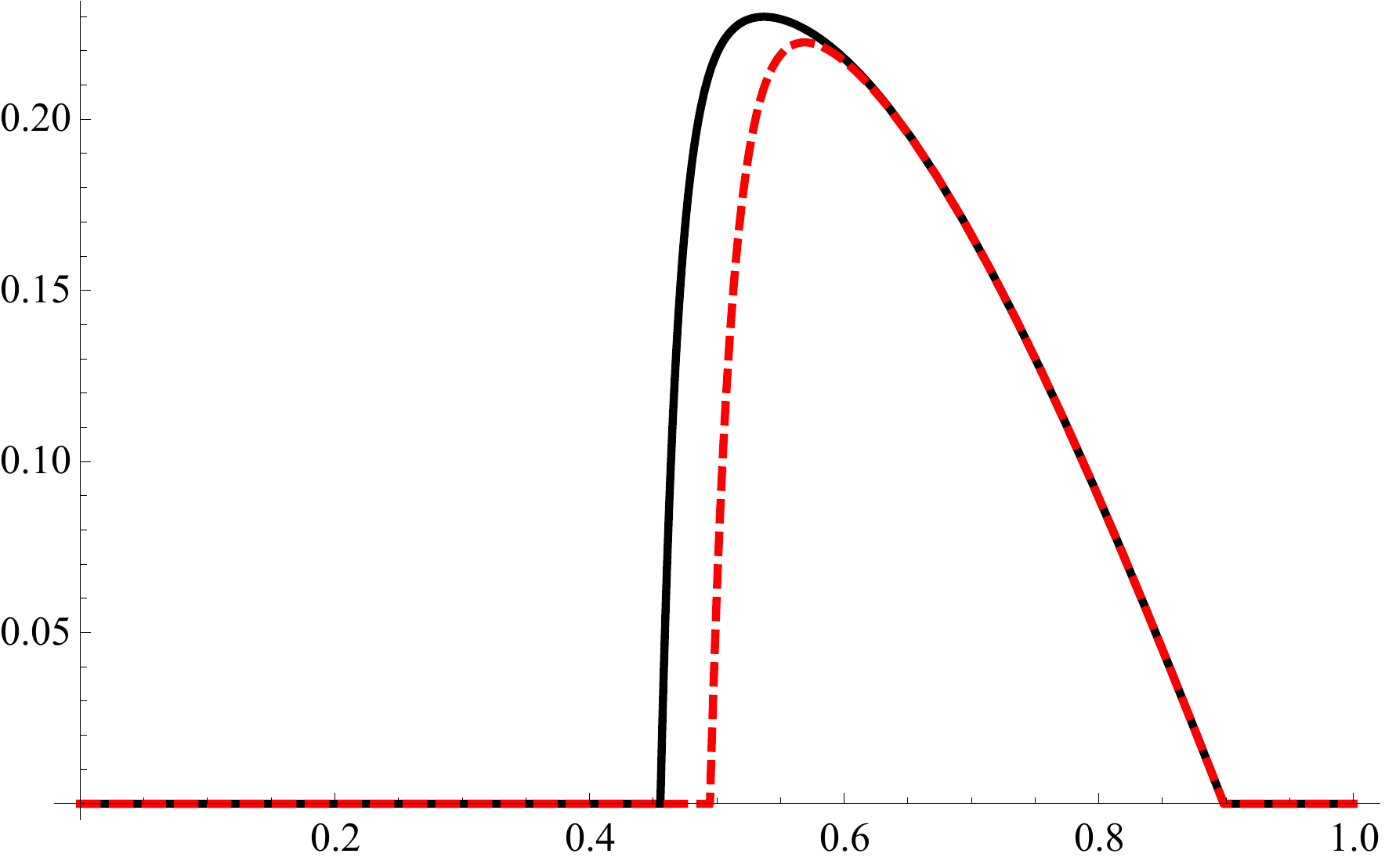}\label{fig02a}}
\subfigure[$A_2$: $\Gamma_1$ (black, solid), $\Gamma_2$ (red, dashed)]
{\includegraphics[height=0.23\textheight,width=0.475\textwidth]{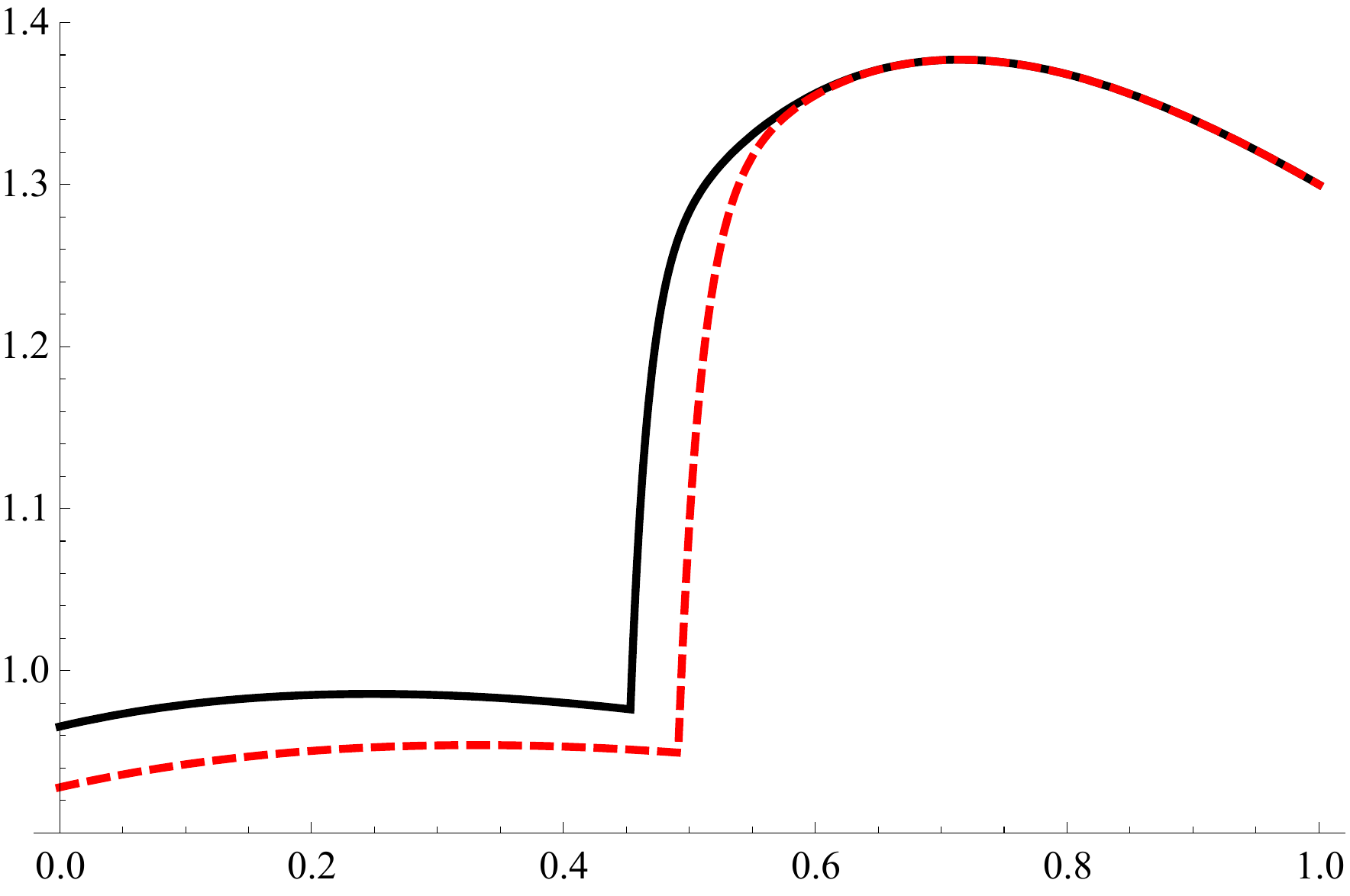}\label{fig02b}}
\caption{Expected payoffs of hackers versus the transformed ransom amount  $(r+1)^{-1}$.}
\end{figure}
\end{Exa}

What if the condition $c_3\le p_3c_1$ in Proposition \ref{pro06} does not hold?  That is, the  recovery cost is relatively large. We show in the following example  that when $c_3>p_3c_1$, both types of hackers may have  larger expected payoffs in   game $\Gamma_2$ than that in   game $\Gamma_1$ in equilibrium.

\begin{Exa}[Example \ref{exa01} continued]\rm
Consider two games $\Gamma_1$ and $\Gamma_2$ with the same setting in Example \ref{exa01}. {Here, we assume $c_1=0.3$, $c_3=0.6$, $p_1=0.3$ and $p_3=0.3$. Clearly, $c_3=0.6>0.3\times 0.3=p_3\times c_1$,} violating the condition in Proposition \ref{pro06}. Figures \ref{fig03a} and \ref{fig03b} plot the curves of  type $A_1$ and type $A_2$ hacker's expected payoffs in  games $\Gamma_1$ and $\Gamma_2$, respectively. {It is seen that both types of hackers can achieve larger maximum expected payoffs in game $\Gamma_2$.}

\begin{figure}[!htbp]\centering
\subfigure[$A_1$: $\Gamma_1$ (black, solid), $\Gamma_2$ (red, dashed)]
{\includegraphics[height=0.2\textheight,width=0.475\textwidth]{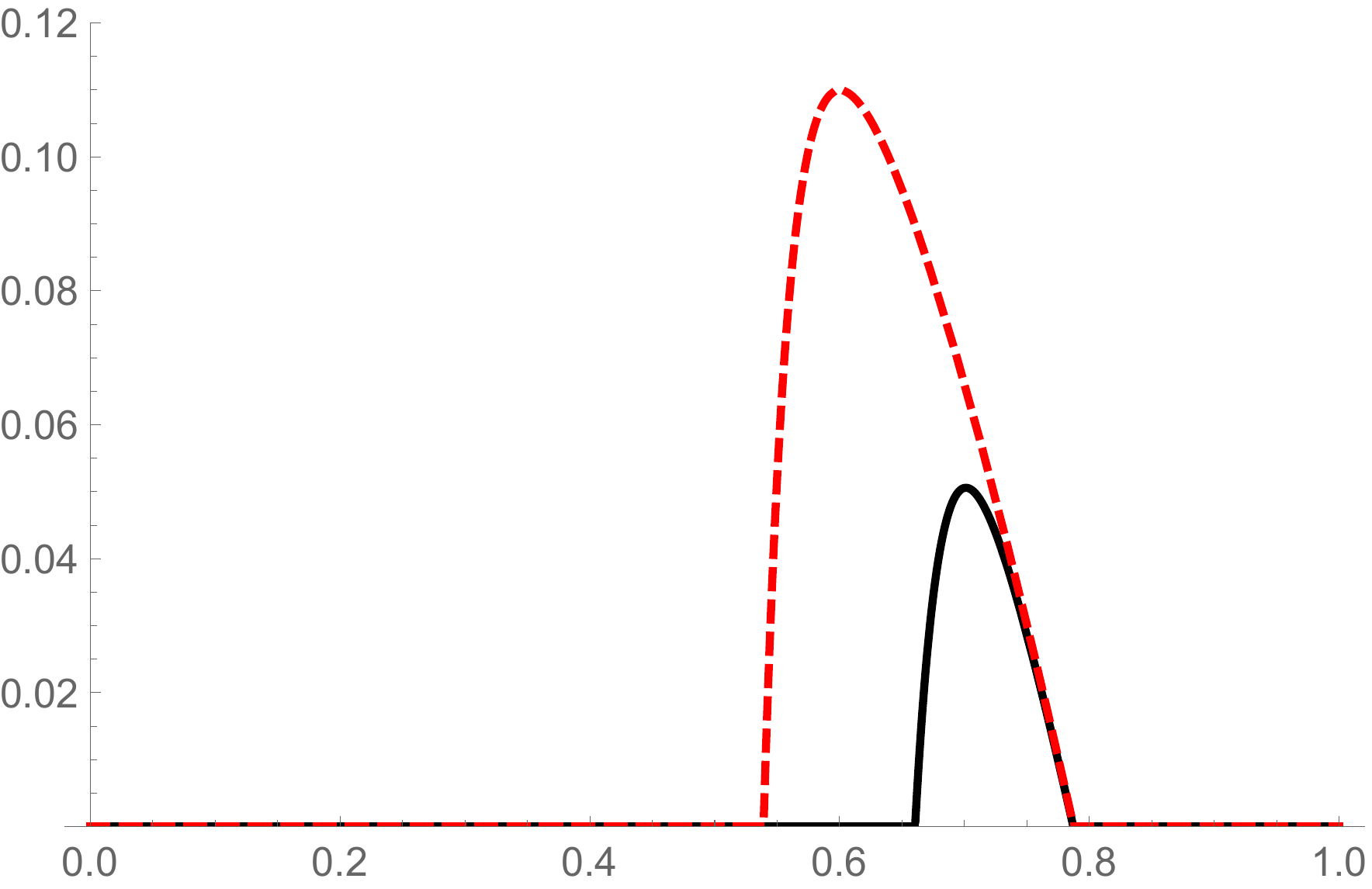}\label{fig03a}}
\subfigure[$A_2$: $\Gamma_1$ (black, solid), $\Gamma_2$ (red, dashed)]
{\includegraphics[height=0.23\textheight,width=0.475\textwidth]{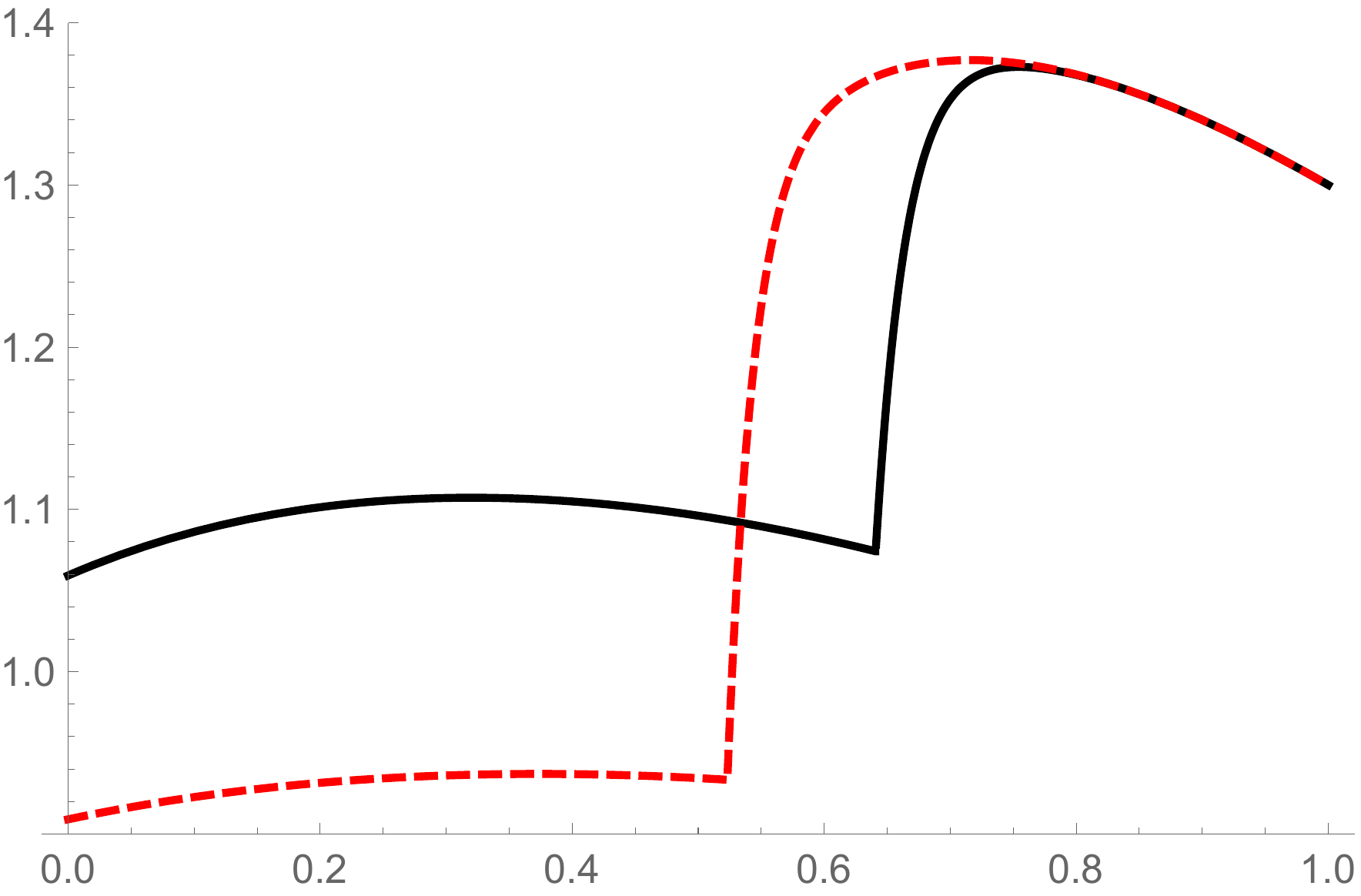}\label{fig03b}}
\caption{Expected payoffs of hackers versus the transformed ransom amount  $(r+1)^{-1}$.}
\end{figure}
\end{Exa}

Typically, the smaller the expected payoff is, the less incentive of the hacker will be. As illustrated in the above examples, employing backups may not necessarily lower the hacker's expected payoffs, when the cost of recovery is relatively high comparing to the cost of cracking. In fact, if the recovery cost is high enough, adding the backup strategy may stimulate the hacker to launch the attack.

\section{Concluding remarks}
In this paper, we propose Bayesian game frameworks with incomplete information to model the conflicting situation between the hacker and the victim in the ransomware ecosystem. In one game, the victim can choose from discarding, paying or cracking; in the other game, the victim has an additional option of recovering from backups. Bayesian Nash equilibria of pure and randomized strategies for each game are provided. The effect of different parameters on the hacker's expected payoffs is also discussed. Our finds can answer the following interesting questions:
\begin{itemize}
\item[a)] {\em Would the backups be always beneficial?}   Our results show that backups could be beneficial or harmful to the victim's combat against ransomware, depending on its cost comparing to the cost of cracking.
\item[b)] {\em Should  the ransom not be paid as suggested by the government authorities?}  According to our study, when the ransom amount is not very large compared to the valuation of files, the victim should pay the ransom directly. However, when the ransom is large, the victim should never pay the ransom directly.
\item[c)] {\em  Which type of hackers can make more money?}  We found that type $A_2$ hacker can always make more money than type $A_1$ does no matter what the corresponding equilibrium ransom amounts are.  Although the proportion of fake ransomware hackers could be relatively small, due to the ability of obtaining a larger expected payoff, attacks launched by fake ransomware hackers may be more than expected.
\item[d)] {\em  Is it bad to have more type $A_1$ hackers?}  We show that the expected payoffs of both types of hackers increase with the chance of encountering type $A_1$ hacker.
\item[e)] {{\em Are the victims more prone to paying the ransom or discarding the files when facing the type $A_2$ hacker?} Through several numerical examples, it is observed that when facing a type $A_2$ hacker, fewer victims may choose to discard the files and more victims may choose to pay the ransom.}
\end{itemize}

One important characteristic regarding the ransomware games should be pointed out: victims are unable to identify the hacker type. This is caused by several factors: (i) the game is one-shot; (ii) the victims know all the information needed to make an action, but is unable to know the attacking cost of the hackers and the earning ability of the type $A_2$ hacker; (iii) the relationship between the equilibrium ransom amounts of the two types of hackers is indefinite. These factors make it challenge for the victims to identify the hacker type with only the information of ransom amount.

The game models studied in this paper can be extended in several directions. Due to the economic and risk characteristics of ransomware,  people should seek more strategies in combating with hackers.  One important strategy is to  enhance the education  of cybersecurity in the organization, which would improve the safety level and efficiently prevent the ransomware attacks.  The other potential option is to purchase the cybersecurity insurance to cover a part  or the total amount of ransom, which will mitigate the economic loss caused by paying the ransom.  However, given the immature status of  cybersecurity insurance market \cite{eling2018cyber}, this study will be pursued in a future work.


 \bibliographystyle{ieeetr}

\bibliography{rans}

\end{document}